\documentclass[prb,floatfix,aps,twocolumn,showpacs,superscriptaddress]{revtex4}
\usepackage[dvips]{graphics}
\usepackage[english]{babel}
\usepackage{graphicx}
\usepackage{amssymb}
\usepackage{epstopdf}
\usepackage{latexsym}
\usepackage{amsmath}
\usepackage{color}

\begin{document}
\title{Free Energy Analysis of Spin Models on Hyperbolic Lattice Geometries}

\author{Marcel Serina}
\affiliation{Department of Physics, University of Basel, Klingelbergstrasse 82,
CH-4056 Basel, Switzerland}
\author{Jozef Genzor}
\affiliation{Institute of Physics, Slovak Academy of Sciences, D\'{u}bravsk\'{a} cesta 9, SK-845 11, Bratislava, Slovakia}
\author{Yoju Lee}
\affiliation{Faculty of Physics, University of Vienna, Boltzmanngasse 5, A-1090 Vienna, Austria}
\author{Andrej Gendiar}
\affiliation{Institute of Physics, Slovak Academy of Sciences, D\'{u}bravsk\'{a} cesta 9, SK-845 11, Bratislava, Slovakia}

\date{\today}

\begin{abstract}
We investigate relations between spatial properties of the free energy and the radius
of Gaussian curvature of the underlying curved lattice geometries. For this purpose we
derive recurrence relations for the analysis of the free energy normalized per lattice
site of various multistate spin models in the thermal equilibrium on distinct
non-Euclidean surface lattices of the infinite sizes. Whereas the free energy is
calculated numerically by means of the Corner Transfer Matrix Renormalization Group
algorithm, the radius of curvature has an analytic expression. Two tasks are considered
in this work. First, we search for such a lattice geometry, which minimizes the free
energy per site. We conjecture that the only Euclidean flat geometry results in the
minimal free energy per site regardless of the spin model. Second, the relations among
the free energy, the radius of curvature, and the phase transition temperatures are
analyzed. We found out that both the free energy and the phase transition temperature
inherit the structure of the lattice geometry and asymptotically approach the profile
of the Gaussian radius of curvature. This achievement opens new
perspectives in the AdS-CFT correspondence theories.
\end{abstract}


\maketitle

\section{Introduction}

The thermodynamic properties and the phase transition phenomena of various physical
systems on two-dimensional non-Euclidean surfaces have attracted the attention of
many theorists and experimentalists for a couple of decades. Especially, the studies
of the hyperbolic surfaces, i.e. the negatively curved geometry, exhibit the increasing
interest in theoretical research of quantum gravity, where the anti-de Sitter (AdS)
hyperbolic spatial geometry plays its essential role. Thus, the mutual interplay among
condensed matter physics, the general theory of relativity, and conformal field theory
(CFT) enriches the interdisciplinary research~\cite{AdS-CFT}. Among them, let us mention
experiments on magnetic nanostructures~\cite{exp1,exp2,exp3}, soft materials of
a conical geometry~\cite{conic}, lattice dislocations of the solid-state crystals,
quantum gravity~\cite{QG1,QG2}, and complex networks~\cite{CN1,CN2}, where neural networks
with the non-Euclidean geometry belong to as well.

A typical theoretical example of such a hyperbolic surface geometry is a two-dimensional
discretized hyperbolic lattice with a constant negative Gaussian curvature. We consider an
infinite set of such hyperbolic lattices constructed by the regular tessellation of
congruent polygons, which are connected without empty spaces at the lattice sites (the
vertices) with fixed coordination numbers. The hyperbolic lattices of finite sizes are
known for their enormous number of the boundary sites. The number of the boundary
sites is always larger than the remaining number of all the inner sites. Or, equivalently,
if one gradually increases the size of a hyperbolic lattice by regularly adding the
outermost layers, the total number of the (boundary) sites increases exponentially with
respect to the increasing radius of the lattice. Since we also intend to determine the
phase transitions of various multi-spin Hamiltonians on the hyperbolic lattices, an
accurate numerical algorithm has to be used, which is also capable of treating the models
in the thermodynamic limit, where the perimeter of the lattice (the size) is infinite.
Such a condition makes the spin systems extremely difficult to be treated numerically
on the hyperbolic lattice geometries (in general, these systems are not integrable).
Therefore, the transfer matrix diagonalization methods are not numerically feasible
(due to a non-trivial way of the transfer matrix construction) and the Monte Carlo
simulations are not completely reliable (due to the insufficiency of the finite-size scaling
near phase transitions).

We have recently proposed an algorithm~\cite{hCTMRG54}, which generalized the
Corner Transfer Matrix Renormalization Group (CTMRG) method~\cite{SqCTMRG}. So far,
we applied the method to study Ising-like systems on certain types of the hyperbolic
lattices, where either the lattice coordination number was fixed letting the polygons
to vary~\cite{hCTMRGp4} or we fixed the polygons to be triangles and varied the
coordination number~\cite{hCTMRG3q}.
In the present work we expand our earlier studies to multi-state spin Hamiltonians on a
much broader set of the hyperbolic lattices so that both the coordination number $q$ and
the number of the sides $p$ in the polygons can vary. We describe a unique way of deriving
generalized recurrence relations for such lattices by the CTMRG algorithm, which
enable us to study phase transitions of the $M$-state clock and $M$-state Potts spin models
($M\geq2$) on any lattice geometries for arbitrary polygon number $p\geq4$ and for an
independent coordination number $q\geq4$.

Particular attention is focused on the analytic derivation of the free energy per spin
site via the calculation of the normalized partition function by CTMRG. The free energy
per site is a well-conditioned thermodynamic function, which does not diverge in the
thermodynamic limit. The numerical calculation of the free energy by CTMRG reaches a
high accuracy; as it will be evident from singular behavior of the specific heat at a
phase transition even after taking the second derivative of the free energy with respect
to temperature numerically. (Notice that Monte Carlo simulations are inefficient in
evaluating the free energy due to large numerical fluctuations). The free-energy analysis
has never been considered in any non-Euclidean systems yet.

Hence, the current numerical analysis may serve as an appropriate, accurate, as well as
complementary source of information for the non-integrable spin systems. The determination
of the phase transition point can be derived, for instance, from the specific heat, which
exhibits non-analytic behavior at a phase transition. The free energy naturally contains an
important information, which reflects a rich boundary structure of the underlying lattice.
In our previous studies, we mainly used to analyze single-site expectation values, such as
the spontaneous magnetization (measured in the lattice center only). In that case the
boundary effects are always negligibly small, and the results are in full agreement with
the known exact solution for the Ising model on the Bethe lattice. As we show later,
having considered the boundary effects, the phase transition is completely suppressed
and can be restored by an appropriate redefinition of the free energy. The correctness
of the present numerical calculations is compared with the exact solutions of the phase
transition in the Ising model on various types of Bethe lattices~\cite{Baxter}.

One can reverse the order of considerations and put another non-trivial question, which
we intend to answer in the current study. The question is associated with a particular
interest in the AdS-CFT correspondence, the so-called gauge duality~\cite{AdS-CFT}.
A highly complicated boundary structure of a finite hyperbolic anti-de Sitter space
(locally viewed as a Minkovski-like space) can be regarded as a spacetime for the conformal
field theory (being identical to a gravitational theory). Our work is focused on the features
of the complex boundary structures only, and no time evolution has been considered. A simple
physical model with a spin-spin interaction network can be used to form a regular hyperbolic
(AdS) space, which can be analyzed with sufficient accuracy. So far, there has been neither
theoretical nor numerical study aimed for the free-energy analysis of the AdS spaces.
However, we have succeeded and gradually developed a way of the free-energy analysis of
such non-Euclidean systems. A condensed-matter point of view on the AdS-CFT correspondence
can undergo difficulties, one of them being the problem of a preferred coordinate system,
i.e., a lattice~\cite{PWA}. For simplicity, we have chosen an infinite set of two-dimensional
curved hyperbolic surfaces (AdS spaces), where the underlying lattice geometry is not
fixed at all, but can vary by changing two integer lattice parameters $p$ and $q$. We
intend to investigate how the ($p,q$) geometry impacts on the total free energy (including
the phase transitions) so that the boundary effects are fully incorporated in the process.

Another question is related to a more concrete physical problem, where we consider an
$M$-state spin Hamiltonian defined on all possible infinite-sized lattice geometries
($p,q$). The spin network is formed in such a way that we allow each multi-state spin
to interact with $q$ nearest-neighboring spins only with a constant interaction coupling
set to the unity. It results in hyperbolic geometries of various Gaussian curvatures.
The free-energy study of the multi-state spin systems for the classical spin system can
be also generalized to the ground-state energy study for the quantum spin system, as we
mention later. We also answer the question, of which of the ($p,q$) lattice geometries
can minimize the free energy per site.

The paper is organized as follows. In Sec.~II we briefly define the regular ($p,q$)
lattice geometries. In the Sec.~II A., the multi-state spin clock and Potts
Hamiltonians are defined. A concise graphical representation of the recurrence
relations is derived in the Sec.~II B (this part can be skipped as it explains
an algorithmic structure of the corner transfer tensors in details). Having derived
the recurrence relations, we check the correctness of the CTMRG algorithm in Sec.~III
by calculating the phase transitions of the simple Ising model on sequences of selected
($p,q$) lattices. The entire Sec.~III can be also skipped since it is devoted to
comparing numerical accuracy of the developed algorithm at phase transitions with
the exactly solvable cases on the Bethe lattices. The analytic derivation of the
free energy per site, which is the core of this study, is given in Sec.~IV. The
numerical results are presented in Sec.~V, where the phase transitions of the multi-state
spin models on the ($p,q$) lattice geometries are analyzed by the free energy per site
and the related specific heat. The concept of the {\it bulk} free energy is defined
in order to extract the correct information on the phase transition, provided that the
boundaries effects are suppressed. The two-dimensional surface profiles of the free energy
with respect to the geometry parameters $p$ and $q$ are calculated. The main purpose is
to study relations between the surface profiles of the free energy and the radius of the
Gaussian curvature in the asymptotic ($p,q$) limit. We observe that the free energy per
site can asymptotically reflects the geometrical structure of the spin-spin interactions
being associated with the underlying lattice geometry. Moreover, we show that the phase
transition temperatures can also copy the ($p,q$) lattice geometry as the free energy does.
In Sec.~VI we discuss our results.

\section{Model and Method}

The idea of replacing the standard transfer matrix formulation of classical spin
systems by the alternative corner transfer matrix method originates in Baxter's
proposal of treating spin Hamiltonians~\cite{Baxter}. The reformulation of Baxter's
analytical study into the numerical CTMRG algorithm was first performed by Nishino
and Okunishi~\cite{SqCTMRG}, who combined the corner transfer matrix formalism with the
numerically effective Density Matrix Renormalization Group method~\cite{White}.
In 2007, the CTMRG algorithm was generalized and applied to the Ising model on the
pentagonal hyperbolic lattice with the constant coordination number four~\cite{hCTMRG54}.

The essence of the CTMRG algorithm rests in finding the recurrence relations, which are
used for the extension of the corner transfer matrices. Before we propose a unified
CTMRG algorithm for any classical spin system on the hyperbolic lattice surfaces, we
describe the lattice geometry that is gradually built up by polygons. Let the lattice
be made by the regular polygonal tessellation with the constant coordination number.
Each lattice geometry is characterized by the Schl\"{a}fli symbol ($p,q$), where $p$
is associated with the regular polygon of $p$ sides (the $p$-gon in the following)
with the constant coordination number $q$.

There are three possible scenarios of creating the lattice geometry ($p,q$) for
the integers $p>2$ and $q>2$. (1) The condition $(p-2)(q-2)=4$ gives rise to the 
two-dimensional Euclidean flat geometry. In this study, we consider only the square
lattice ($4,4$), which satisfies the condition, and the remaining triangular ($3,6$)
and honeycomb ($6,3$) Euclidean lattices will be studied elsewhere. (2) If $(p-2)(q-2)>4$,
the infinite set of the hyperbolic geometries can satisfy the condition. Although 
such lattices of infinite size define various two-dimensional curved surfaces, the entire
infinite hyperbolic lattice can be spanned in the infinite-dimensional space only; it
it commonly associated with the Hausdorff dimension, which is infinite. None of the
hyperbolic lattices can be endowed in the three-dimensional space. (3) The condition
$(p-2)(q-2)<4$ corresponds to only five finite-sized spherically curved geometries,
which are trivial and are not considered in the current study.

\subsection{The Lattice Model}

Each vertex of the infinite ($p,q$) lattice, built up by the $p$-gons with the
fixed coordination number $q$, represents a classical multi-spin variable $\sigma$
interacting with the $q$ nearest-neighboring spins. The Hamiltonian ${\cal H}_{(p,q)}$
can be decomposed into the sum of identical local Hamiltonians ${\cal H}_p$ acting
exclusively on the local $p$-gons, which are considered to be the basic elements in
the construction of the entire lattice. In particular, the decomposition of the
full Hamiltonian is
\begin{equation}
{\cal H}_{(p,q)}\{\sigma\} = \sum\limits_{(p,q)} {\cal H}_p[\sigma],
\end{equation}
where the sum is taken through the given lattice geometry ($p,q$) accordingly.
The simplified spin notations $[\sigma]$ and $\{\sigma\}$, respectively, are ascribed
to the $p$ spins within each local Hamiltonian ${\cal H}_p[\sigma]\equiv{\cal H}_p
(\sigma_1\sigma_2\cdots\sigma_p)$ and the infinitely many spins $\{\sigma\}$ of the
entire system ${\cal H}_{(p,q)}\{\sigma\}\equiv{\cal H}_{(p,q)}(\sigma_1\sigma_2\cdots
\sigma_{\infty})$. We consider two types of the multi-state
spin models: the $M$-state clock model with the local Hamiltonian
\begin{equation}
{\cal H}_p[\sigma] = -J\sum\limits_{i=1}^{p}
\cos\left[\frac{2\pi}{M}(\sigma_{i}-\sigma_{i+1})\right]
\end{equation}
and the $M$-state Potts model
\begin{equation}
{\cal H}_p[\sigma] = -J\sum\limits_{i=1}^{p} \delta_{\sigma_{i},\sigma_{i+1}}\, ,
\end{equation}
where $\sigma_{p+1}^{~}\equiv\sigma_{1}^{~}$ within the $p$-gon, and where each $M$-state
spin variables $\sigma=0,1,2,\dots,M-1$. (Thus, the Ising model is associated with $M=2$.)
We consider the ferromagnetic interaction $J$ to avoid frustration.

Let the Boltzmann weight ${\cal W}_{\rm B}[\sigma]=\exp(-{\cal H}_p[\sigma]/k_{\rm B}T)$
be defined on the $p$-gon of the local Hamiltonian, where $k_{\rm B}$ and $T$ correspond
to the Boltzmann constant and temperature, respectively. Dimensionless units are used
by setting $J=k_{\rm B}=1$. In general, the Ising model on the hyperbolic lattices ($p,q$)
is not exactly solvable, except for special asymptotic cases, on the Bethe lattices
(when $p\to\infty$), as discussed and numerically checked in Sec.~III.

We employ the generalized CTMRG algorithm as a powerful numerical tool to study
the phase transitions on the arbitrary lattice geometries ($p,q$). The CTMRG algorithm
is an RG-based iterative numerical method, which makes it possible to evaluate the partition
function ${\cal Z}$ and the thermodynamic functions within high accuracy~\cite{Axelrod}.
Let each CTMRG iteration step be enumerated by an integer variable $k$. At the very
beginning of the iterative process, the lattice size of the ($p,q$) geometry is as
small as the size of $q$ connected $p$-gons around a single central spin,
and is referred to as the first iteration step
with $k=1$. In the second iteration step, $k=2$, the lattice can expand its size, i.e.
the number of the spin sites, either as a power law [only for the Euclidean ($4,4$) lattice]
or the size grows exponentially [for all the remaining ($p,q$) hyperbolic cases]. The
lattice size increases with respect to the number of the Boltzmann weights (or,
equivalently, with respect to the total number of the spin sites). Since we are
are interested in the phase transition studies, the thermodynamic limit requires
to take the thermodynamic limit $k\to\infty$, which is equivalent to the case when
the iterative process proceeds until all of the thermodynamic functions (normalized
to the spin site) converge completely.

\begin{figure}[tb]
{\centering\includegraphics[width=\linewidth]{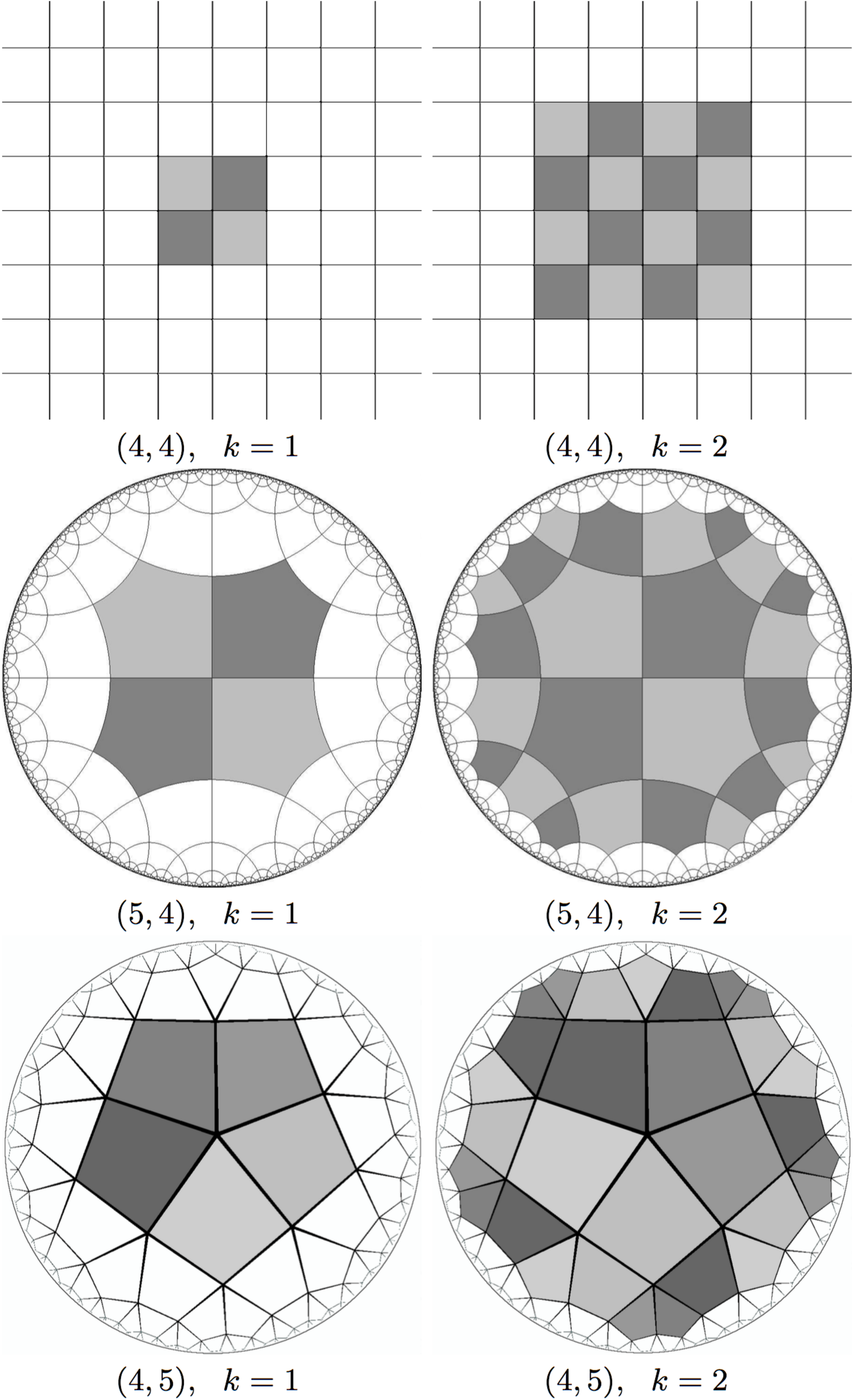}}
  \caption{The illustration of the three selected lattice geometries ($4,4$), ($5,4$), and
($4,5$). The first two CTMRG iteration steps $k=1$ (left) and $k=2$ (right) show the building
process of the lattices by means of the $p$-gonal Boltzmann weight tessellation with the
uniform coordination number $q$. The Boltzmann weights for given $k$ are represented by
the shaded regular (congruent) $p$-gons.}
  \label{fig:01}
\end{figure}

\subsection{Recurrence Relations}

By the end of Sec.~II, we derive the recurrence relations, which are required for
the construction of the CTMRG algorithm. The consequent Sec.~III serves as a benchmark,
where the Ising model on ($p,q$) lattices is treated by CTMRG and its high numerical
accuracy is confirmed by calculating the analytically solvable Ising model on the Bethe
($\infty,q$) lattices. Finally, the three subsections in Sec.~IV systematically build up
a general formula of the free energy for an arbitrary ($p,q$) geometry. All of these
three sections briefly summarize the algorithmic essence of CTMRG, which is necessary
for understanding its mathematical concept. Thus, the readers can freely omit these
sections and proceed from Sec.~V.

The complete expansion (iteration) process is given by recurrence relations as we
specify below. For the more instructive understanding, the derivation of the
recurrence relations is structured into the following three steps:
\begin{itemize}
\item[(i)] $(4,4)$, $(5,4)$, and $(4,5)$,
\item[(ii)] $(4,4)\to(5,4)\to(6,4)\to\cdots\to(\infty,4)$,
\item[(iii)] $(p,q)$.
\end{itemize}

Figure~\ref{fig:01} depicts three typical lattices in the first two iteration steps
($k=1$ and $k=2$). The shaded $p$-gons represent the corresponding finite lattice made
of the Boltzmann weights ${\cal W}_{\rm B}$ at given $k$. The surrounded $p$-gons
shown in white color around the shaded ones stand for the consequent iteration steps. 
The spin variables $\sigma$ are positioned on the vertices of the polygons,
and the sides of the $p$-gons correspond to the constant nearest-neighbor spin
coupling $J=1$. Notice that the sizes and the shapes of the polygons are kept equal
for each lattice geometry ($p,q$), and we display each hyperbolic lattice geometry in
the Poincare disk representation~\cite{Poincare}, which projects the entire hyperbolic
lattice onto the shown unitary circles. As the consequence of that projection, the
sizes of the $p$-gons get deformed and shrunk from the lattice center toward the
circumference of the circle. The circumference is associated with the lattice boundary
in the infinity.

(i) The iterative expansion process is formulated in terms of the generalized corner
transfer matrix notation (for details, see Refs.~\cite{hCTMRG54,hCTMRGp4,hCTMRG3q,
hCTMRG3qn}), where the corner transfer tensors ${\cal C}_j$ and the transfer tensors
${\cal T}_j$ expand their sizes as the iteration step (indexed by $j$) increases, i.e.,
$j=1,2,3,\dots,k$
\begin{equation}
\left.
\begin{split}
{\cal C}_{j+1}&={\cal W}_{\rm B} {\cal T}_{j}^{2} {\cal C}_{j}^{~}\\
{\cal T}_{j+1}&={\cal W}_{\rm B} {\cal T}_{j}^{~}
\end{split}
\ \ \right\} \ \ {\rm for}\ \ (4,4),
\label{rr44}
\end{equation}
\begin{equation}
\left.
\begin{split}
{\cal C}_{j+1}&={\cal W}_{\rm B} {\cal T}_{j}^{3} {\cal C}_{j}^{2}\\
{\cal T}_{j+1}&={\cal W}_{\rm B} {\cal T}_{j}^{2} {\cal C}_{j}^{~}
\end{split}
\ \ \right\} \ \ {\rm for}\ \ (5,4),
\label{rr54}
\end{equation}
\begin{equation}
\left.
\begin{split}
{\cal C}_{j+1}&={\cal W}_{\rm B} {\cal T}_{j}^{2} {\cal C}_{j}^{3}\\
{\cal T}_{j+1}&={\cal W}_{\rm B} {\cal T}_{j}^{~} {\cal C}_{j}^{~}
\end{split}
\ \ \right\} \ \ {\rm for}\ \ (4,5).
\label{rr45}
\end{equation}
The tensors are initialized to the Boltzmann weight and
${\cal C}_{1}^{~}={\cal T}_{1}^{~}\equiv {\cal W}_{\rm B}$.

The recurrence relations in Eqs.~\eqref{rr44}-\eqref{rr45} are written in a simplified form,
it means we excluded the indexing of the lattice geometry ($p,q$) they depend on. Hence,
we omit the lattice superscript so that ${\cal C}_{j}^{(p,q)} \to {\cal C}_{j}^{~}$ and
${\cal T}_{j}^{(p,q)} \to{\cal T}_{j}^{~}$. The partition function, ${\cal Z}_{(p,q)}^{[k]}$,
in the final $k^{\rm th}$ iteration step is given by the configuration sum (or, equivalently,
by the trace) of the product of the $q$ corner transfer tensors, which are concentrically
connected around the central spin site of the lattice~\cite{hCTMRG54}
\begin{equation}
{\cal Z}_{(p,q)}^{[k]} = {\rm Tr}\left[e^{-{\cal H}_{(p,q)}/T}\right]
={\rm Tr}\ (\underbrace{{\cal C}_{k}{\cal C}_{k}\cdots{\cal C}_{k}}_{q})
\equiv{\rm Tr}\ {\left({\cal C}_{k}\right)}^{q}.
\label{part_fnc}
\end{equation}
The evaluation of the partition function via the product of the Boltzmann weights of the
$p$-gonal shape can be also expressed graphically, which may serve as a visual simplification
of Eq.~\eqref{part_fnc}. For instance, the size of the square lattice ($4,4$) in the second
iteration step, $k=2$, corresponds to the evaluation of the partition function
${\cal Z}_{(4,4)}^{[k=2]}$. This is equivalent to the product of the $16$ Boltzmann
weights in accord with the respective lattice shown in Fig.~\ref{fig:01}, i.e.,
\begin{equation}
{\cal Z}_{(4,4)}^{[2]}
 = {\rm Tr}\ {\left({\cal C}_{2}\right)}^{4}
 = {\rm Tr}\ {\left({\cal W}_{\rm B}{\cal T}_{1}^{2}{\cal C}_{1}^{~}\right)}^{4}
 = {\rm Tr}\ {\left({\cal W}_{\rm B}\right)}^{16}.
\end{equation}
Thus, the power of ${\cal W}_{\rm B}$ matches the total number of the shaded
squares in Fig.~\ref{fig:01} for given $k$. The number of the square-shaped Boltzmann
weights obeys the power law $4k^2$ being the number of the squares on the ($4,4$)
lattice for given $k$.

The partition functions of the two hyperbolic lattices ($5,4$) and ($4,5$), as selected
within (i), are evaluated analogously. The lattice size in the second iteration step, $k=2$,
is graphically sketched in Fig.~\ref{fig:01} and is related to taking the configuration
sum over the product of the shaded $p$-gons. For the instructive purpose, the two respective
partition functions for $k=2$ satisfy the expressions
\begin{equation}
{\cal Z}_{(5,4)}^{[2]}
 = {\rm Tr}\ {\left({\cal C}_{2}\right)}^{4}
 = {\rm Tr}\ {\left({\cal W}_{\rm B}{\cal T}_{1}^{3}{\cal C}_{1}^{2}\right)}^{4}
 = {\rm Tr}\ {\left({\cal W}_{\rm B}\right)}^{24}
\end{equation}
and
\begin{equation}
{\cal Z}_{(4,5)}^{[2]}
 = {\rm Tr}\ {\left({\cal C}_{2}\right)}^{5}
 = {\rm Tr}\ {\left({\cal W}_{\rm B}{\cal T}_{1}^{2}{\cal C}_{1}^{3}\right)}^{5}
 = {\rm Tr}\ {\left({\cal W}_{\rm B}\right)}^{30}\, ,
\end{equation}
where the powers of ${\cal W}_{\rm B}$ on the right hand side of the equations count
the number of the $p$-gonal Boltzmann weights. We recall that the total number of the
Boltzmann weights grows exponentially with respect to the iteration step $k$. The
analytic formula of the exponential dependence of the total number of the spin sites
on $k$ is derived in the following section, where the free energy is examined in detail.

(ii) We recently investigated the Ising model on an infinite sequence of hyperbolic
lattices~\cite{hCTMRGp4}, for which the coordination number was fixed to $q=4$, whereas
the size of the $p$-gons increased $p=4,5,6,\dots,\infty$. The generalized recurrence
relations satisfying the lattices ($p\geq4,4$) are summarized into a more compact form
\begin{equation}
\begin{split}
{\cal C}_{j+1}& = {\cal W}_{\rm B} {\cal T}_{j}^{p-2} {\cal C}_{j}^{p-3}\, ,\\
{\cal T}_{j+1}& = {\cal W}_{\rm B} {\cal T}_{j}^{p-3} {\cal C}_{j}^{p-4}\, .
\end{split}
\end{equation}
We conjectured~\cite{hCTMRGp4} that the Ising model realized on the sequence of the
lattices $\{(4,4),(5,4),(6,4),\dots,(\infty,4)\}$ converges to the Bethe lattice
with the coordination number $q=4$ exponentially with the increasing $p$. In
other words, the Bethe lattice is actually identical with the lattice geometry
($\infty,4$). We showed, when evaluating all the thermodynamic functions, that
any lattice geometry ($p\geq15,q=4$) was numerically indistinguishable from the Bethe
lattice with high accuracy~\cite{hCTMRGp4}. In particular, having evaluated the phase
transition temperature $T_{\rm pt}^{(\infty,4)}$ of the Ising model on the Bethe lattice,
which had been numerically realized on the ($15,4$) lattice geometry, the numerical
accuracy of the CTMRG algorithm resulted in $T_{\rm pt}=2.88539$. The Ising model on
the Bethe lattice is an exactly solvable system with the phase transition temperature
$T_{\rm pt}=1/\ln\sqrt{2}$ as derived by Baxter~\cite{Baxter}.

(iii) Now we generalize the recurrence relations by considering arbitrary $p$-gons
$p\geq4$ as well as the coordination number $q\geq4$.
Having analyzed all the geometrical lattice structures ($p,q$) of the polygonal
tailing, it straightforwardly leads to the recurrence relations
\begin{equation}
\begin{split}
{\cal C}_{j+1}& = {\cal W}_{\rm B} {\cal T}_{j}^{p-2} {\cal C}_{j}^{(p-2)(q-3)-1},\\
{\cal T}_{j+1}& = {\cal W}_{\rm B} {\cal T}_{j}^{p-3} {\cal C}_{j}^{(p-3)(q-3)-1}.
\end{split}
\end{equation}
The calculation of the partition function ${\cal Z}$ for any ($p,q$) lattice geometry
in the $k^{\rm th}$ iteration step remains identical to Eq.~\eqref{part_fnc}. Therefore,
the expectation value $\langle O\rangle$ of a local observable $O$ is evaluated directly.
The typical example is the spontaneous magnetization $M=\langle\sigma_c\rangle$ measured
in the center of the lattice ($p,q$), where the spin variable $\sigma_c$ is positioned.
If evaluated in the thermodynamic limit, we obtain
\begin{equation}
M_{(p,q)}=\langle\sigma_c\rangle
 =\frac{{\rm Tr}\left[\sigma_c\,e^{-{\cal H}_{(p,q)}/T}\right]}
       {{\rm Tr}\left[          e^{-{\cal H}_{(p,q)}/T}\right]}
=\frac{ {\rm Tr}\,\left[ \sigma_c{\left({\cal C}_{\infty}\right)}^{q}\right] }
      {{\cal Z}_{(p,q)}^{[\infty]}}
\label{Mpq}
\end{equation}
for arbitrary ($p,q$).

\begin{figure}[tb]
{\centering\includegraphics[width=\linewidth]{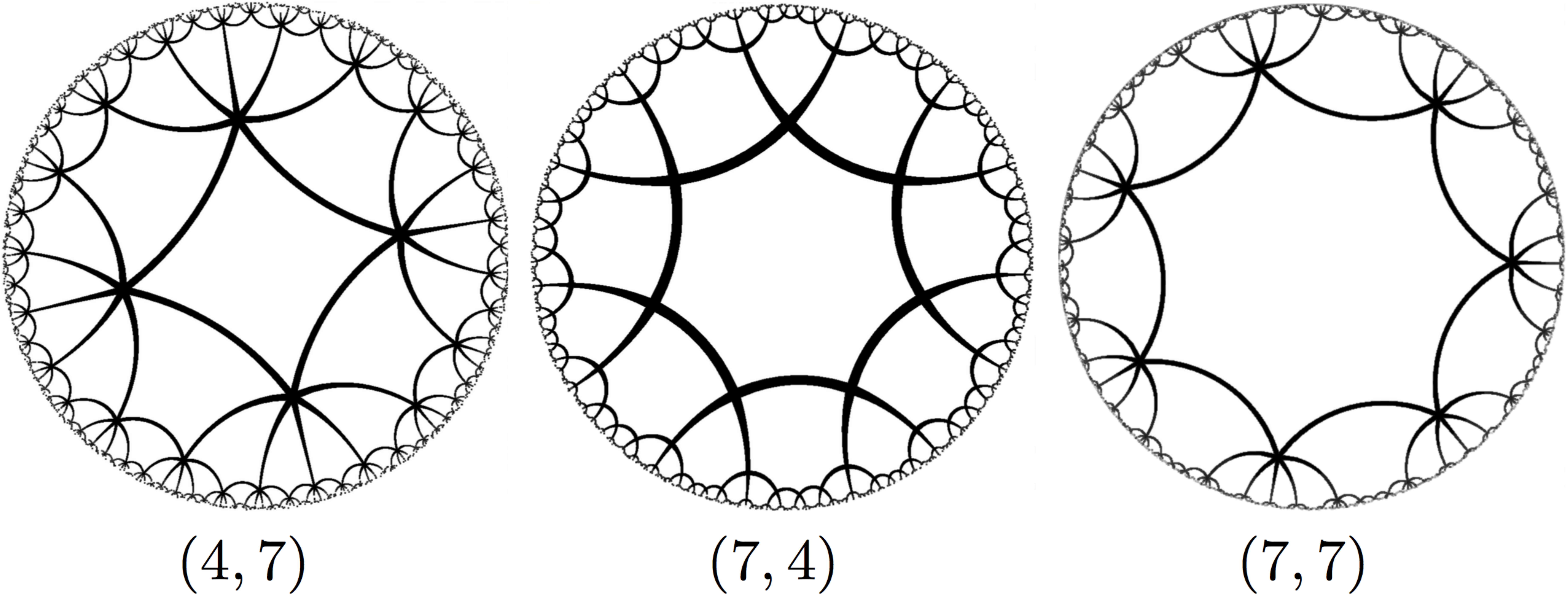}}
  \caption{The Poincar\'{e} disk representation of the three hyperbolic lattices
chosen for the analysis of the thermodynamic functions of the spin model.}
  \label{fig:02}
\end{figure}

\section{Phase Transition Analysis}

\begin{figure}[tb]
{\centering\includegraphics[width=\linewidth]{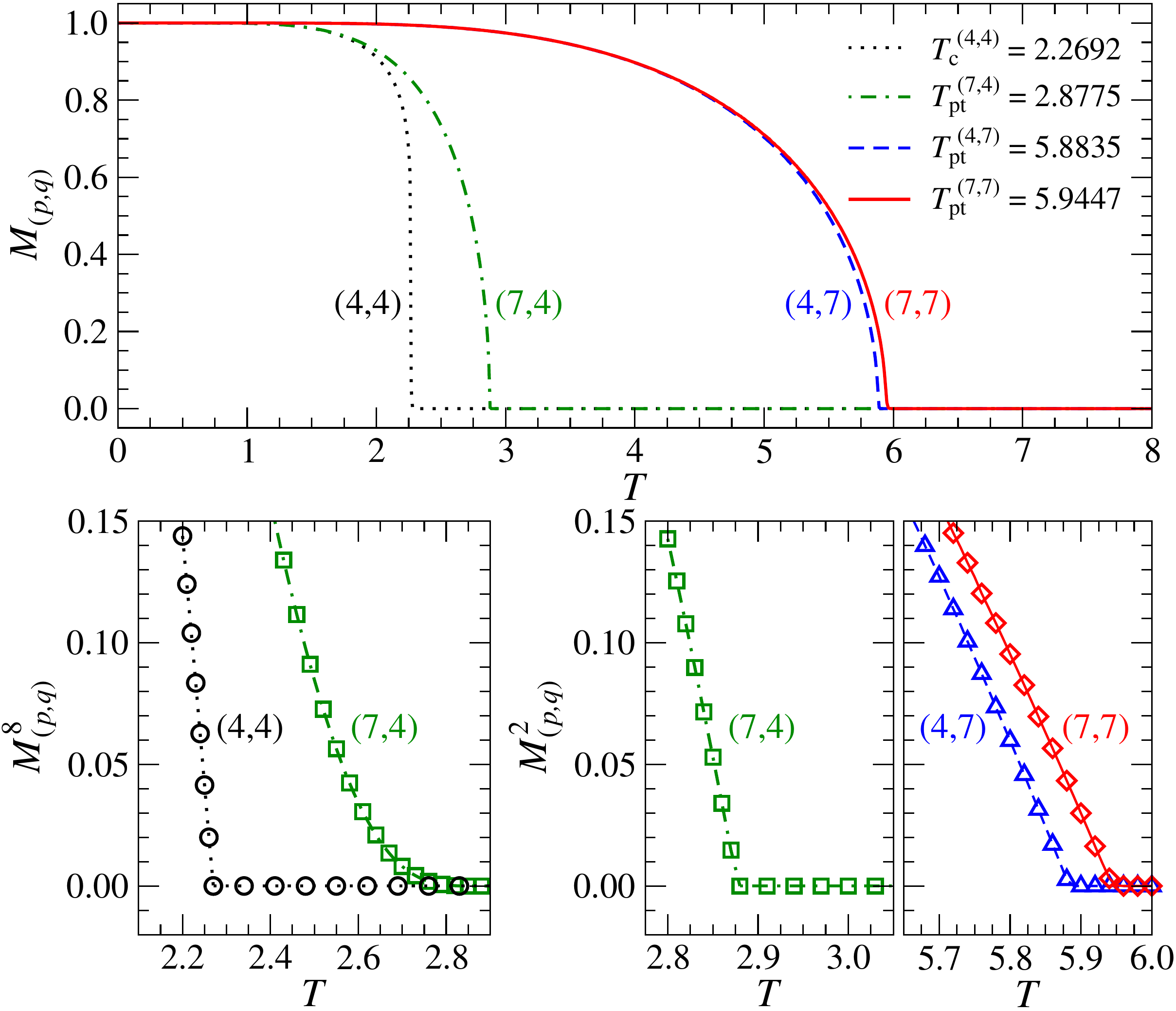}}
  \caption{(Color online) The spontaneous magnetization with respect to temperature for
the Euclidean square lattice as well as for the three hyperbolic lattices depicted in
Fig.~\ref{fig:02}.}
\label{fig:03}
\end{figure}

For instructive reasons, we have selected the three non-trivial hyperbolic lattices ($4,7$),
($7,4$), and ($7,7$), which are shown in Fig.~\ref{fig:02} in the Poincare representation.
We calculated the spontaneous magnetization $M_{(p,q)}$, which is plotted on the top graph
in Fig.~\ref{fig:03} including the case of the Euclidean ($4,4$) lattice, which serves as
as a benchmark since this case is exactly solvable. We have shown~\cite{hCTMRGp4,hCTMRG3q,
hCTMRG3qn} that the Ising model on certain types of the hyperbolic lattices belongs to the
mean-field universality class. Now, we expand our analysis for arbitrary ($p\geq4,q\geq4$)
lattices. The spontaneous magnetization follows the scaling relation $M_{(p,q)}\propto
(T_{\rm pt}^{(p,q)}-T)^{\beta}$ at the phase transition temperature $T_{\rm pt}^{(p,q)}$,
yielding the mean-field magnetic exponent $\beta=\frac{1}{2}$ whenever $(p-2)(q-2)>4$.
Recall that the Ising (not the mean-field) universality class is solely reproduced for
the Ising model on the Euclidean lattices; in this case, it is on the square lattice
($4,4$), where we confirmed that $M_{(4,4)}\propto(T_{\rm pt}^{(4,4)}-T)^{\frac{1}{8}}$.
This is unambiguously manifested by the linear dependence of $M^8_{(4,4)}$ on temperature
$T \leq T_{\rm pt}^{(4,4)}$, as depicted on the lower left graph in Fig.~\ref{fig:03}.

On the other hand, the mean-field universality class with $\beta=\frac{1}{2}$ can be
read off by plotting $M^2_{(p,q)}$ for $T \leq T_{\rm pt}^{(p,q)}$, which is obvious from
the linear decrease of the spontaneous magnetization approaching the phase transition
temperature $T_{\rm pt}^{(p,q)}$, as shown on the two lower graphs on the right side in
Fig.~\ref{fig:03}.

The mean-field-like feature of the spin model is always realized on the hyperbolic lattices.
We point out here that such a mean-field-like behavior is not caused by an insufficient
numerical accuracy. The numerical results are fully converged; any additional increase
of the number of the states kept in the renormalization group algorithm does not improve
any of the thermodynamic functions. Therefore, the reason for the mean-field-like feature
rests in the exceedance of the critical lattice dimension $d_{c}=4$ because the Hausdorff
dimension is infinite for all the hyperbolic lattices in the thermodynamic limit.
The claim is identical to that of the exact solution of the Ising model on the Bethe
lattice, where the analytically derived mean-field exponents on the Bethe lattice have
nothing to do with the mean-field approximation of the model at all~\cite{Baxter}.
Instead, the {mean-field-like} feature is caused by the hyperbolic lattice geometry,
which is accompanied by the absence of the divergent correlation length at the phase
transition~\cite{hCTMRG3q}.

\subsection*{Asymptotic Lattice Geometries}

\begin{figure}[tb]
{\centering\includegraphics[width=\linewidth]{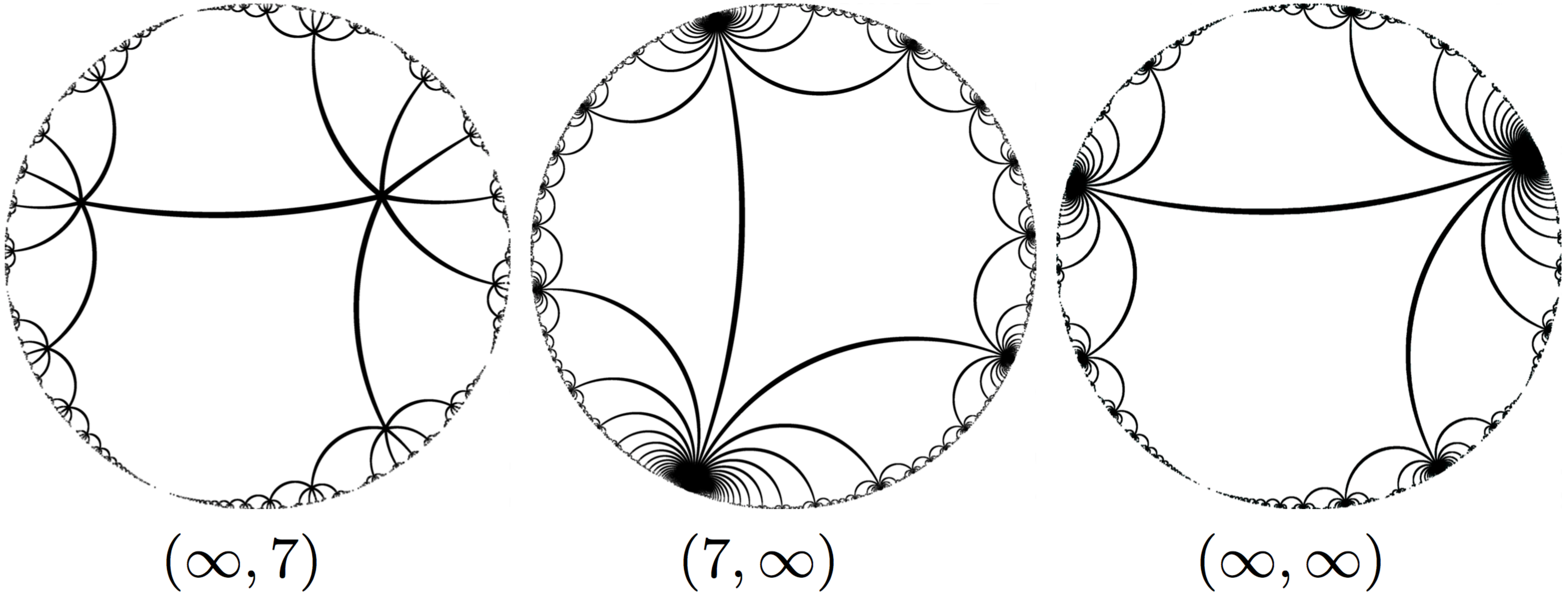}}
  \caption{The Poincar\'{e} representation of the asymptotic hyperbolic lattices
($\infty,7$) on the left, ($7,\infty$) in the middle, and ($\infty,\infty$) on the right.}
\label{fig:04}
\end{figure}

Let us investigate the phase transitions of the Ising model on the {\it asymptotic}
lattice geometries as illustrated in Fig.~\ref{fig:04}. In our earlier studies, we presented
two distinct scenarios: (1) the coordination number was fixed to $q=4$ while the $p$-gons
gradually expanded $p=4,5,6,\dots,\infty$; and (2) we formed the triangular tessellation,
$p=3$, and the coordination number varied $q=6,7,8,\dots,\infty$. In both the cases
a substantially different asymptotic behavior of the phase transition temperatures
was observed~\cite{hCTMRGp4,hCTMRG3q}. In the former case, the phase transition temperature
converges to the Bethe lattice phase transition $T_{\rm pt}^{(p,4)}\to\frac{2}{\ln{2}}$.
In the latter case, the triangular tessellation of the lattice types ($3,q\geq3$) led to
a linear divergence of the phase transition temperature $T_{\rm pt}^{(3,q)}\propto q$.
These findings remain valid for arbitrary ($p,q$) lattices. As examples we selected the
hyperbolic lattices ($7,q$) and ($p,7$) with $p,q=4,5,6,\dots,\infty$ as depicted in
Fig.~\ref{fig:04}.

\begin{figure}[tb]
{\centering\includegraphics[width=\linewidth]{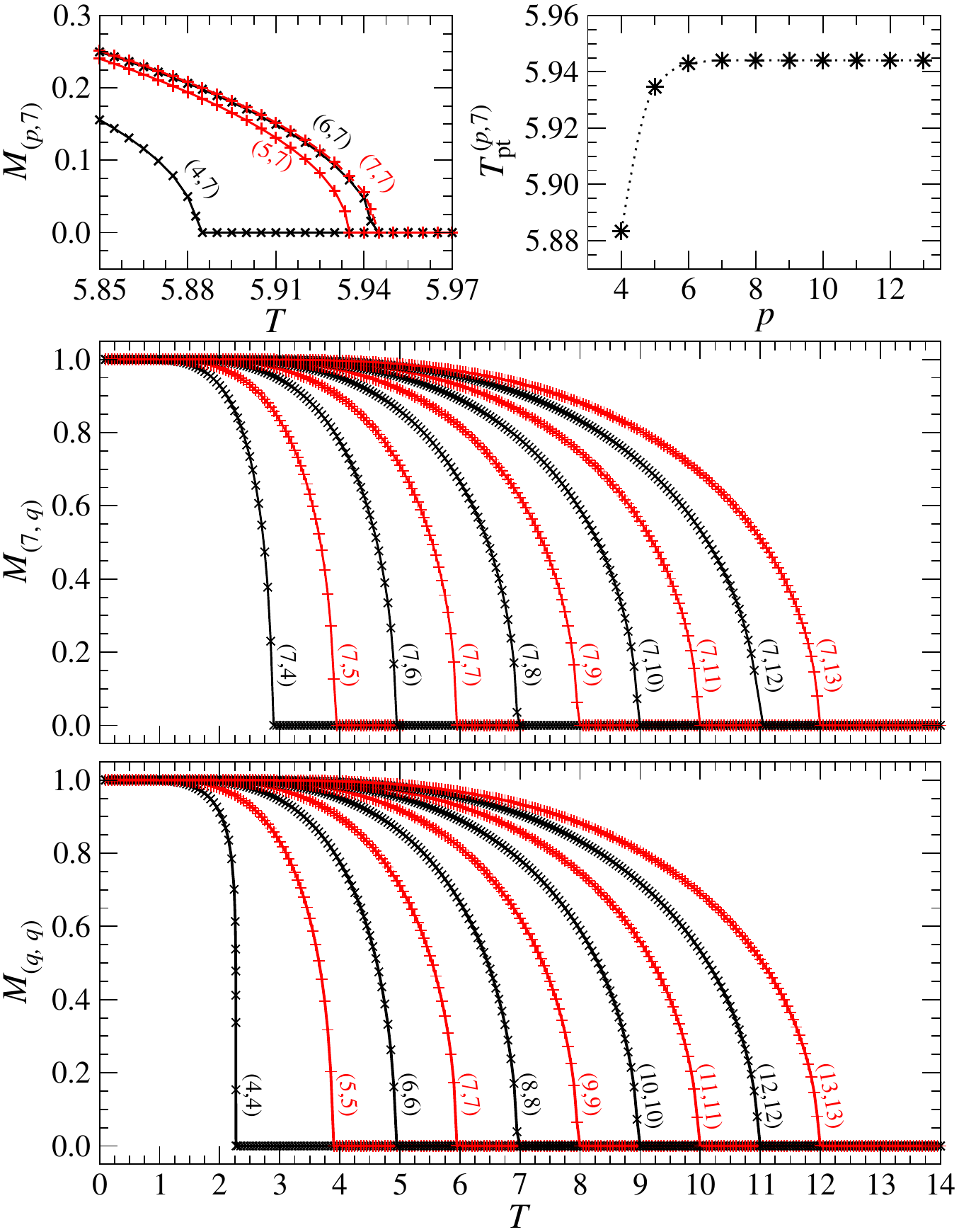}}
  \caption{(Color online) The temperature dependence of the spontaneous magnetization
toward the asymptotic lattice geometries plotted in Fig.~\ref{fig:04}. The insets of the
top panel show the fast convergence of the magnetization, and the model on the lattice
($7,7$) exhibits almost identical behavior as the lattice ($\infty,7$). The top and the
lower graphs, respectively, describe the linear increase of the phase transitions if
the model is studied on the lattices ($7,q$) and ($q,q$) for $q=4,5,6,\dots,\infty$.}
\label{fig:05}
\end{figure}

The two top graphs in Fig.~\ref{fig:05} show a fast convergence of the magnetization
and the phase transition temperatures $T_{\rm pt}^{(p,7)}$ toward the Bethe lattice
($\infty,7$) with the coordination number seven. The fast convergence means that the phase
transition temperature on the ($7,7$) lattice is almost indistinguishable from those on
the consequent ($p>7,7$) lattices. We obtained the asymptotic phase transition temperature
$T_{\rm pt}^{(p\to\infty,7)}\to5.944002$, which is in accurate agreement with the general
formula for the Bethe lattice phase transition temperature~\cite{Baxter},
\begin{equation}
\lim\limits_{p\to\infty}T_{\rm pt}^{(p,q)}=\frac{1}{\ln\sqrt{\left[{q/(q-2)}\right]}}.
\label{Bethe_q}
\end{equation}
The middle graph in Fig.~\ref{fig:05} shows the spontaneous magnetization $M_{(7,q)}$ on
the lattices made from septagonal ($p=7$) tiling for the gradually increasing coordination
number $q=4,5,6,\dots,\infty$. The phase transition temperature diverges linearly. The
result can be generalized, and the linear asymptotic divergence is present,
\begin{equation}
T_{\rm pt}^{(p,q\gg4)}\propto q\, ,
\label{Tpt_q}
\end{equation}
irrespective of $p$. Finally, if both lattice parameters are set to be equivalent,
$p\equiv q$, the scenario with the increasing coordination number and fixed $p$-gon
is dominant over that one with the fixed $q$ and increasing $p$.
The bottom graph in Fig.~\ref{fig:05} depicts the case of the ($q,q$) lattices for
$q=4,5,6,\dots,13$, which also satisfies the linearity $T_{\rm pt}^{(q,q)}\propto q$.

\begin{figure}[tb]
{\centering\includegraphics[width=\linewidth]{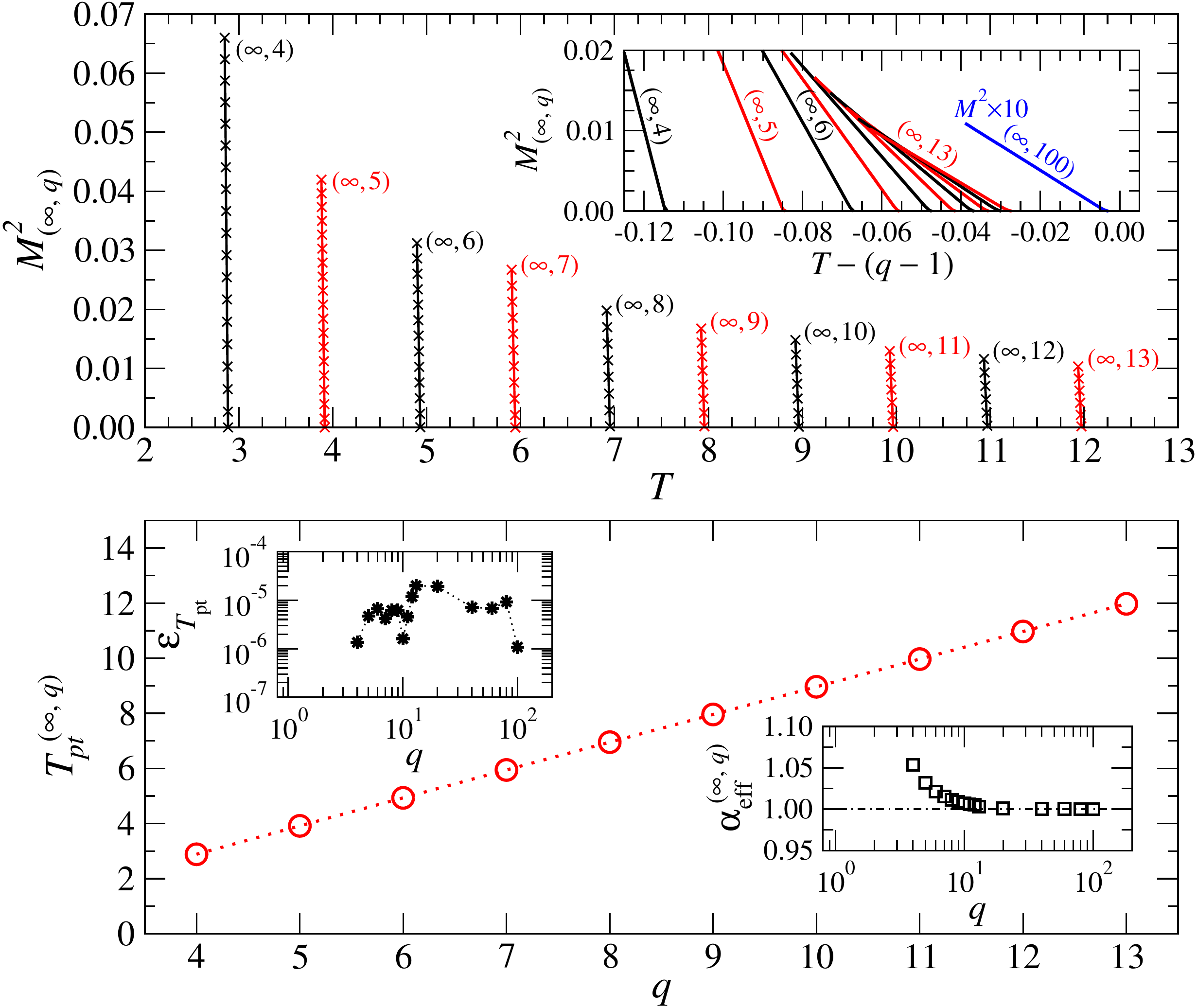}}
  \caption{(Color online) The top graph shows the temperature dependence of a few nonzero
values of the squared magnetization approaching the phase transition from the ordered phase
$T\leq T_{\rm pt}^{(\infty,q)}$ for the lattices $p=20$ and $4\leq q\leq13$ obtained by CTMRG,
which accurately reproduce the Bethe lattice. The inset shows the same data in detail,
rescaled to temperatures $T-(q-1)$. In the lower graph, the linearity of $T_{\rm pt}^{
(\infty,q)}$ on the Bethe lattice is satisfied with increasing $q$. The lower right 
inset shows the convergence of the effective exponent $\alpha_{\rm eff}^{(\infty,q)}\to1$
in the log-lin scale. The top left inset displays the numerical accuracy by evaluating
the relative error for the phase transition temperature on the Bethe lattices.}
\label{fig:06}
\end{figure}

The mean-field universality is induced by the hyperbolic geometry of the curved
two-dimensional surface, which can be spanned only in the infinite-dimensional space
in the thermodynamic limit. In order to examine the asymptotic lattice geometries shown
in Fig.~\ref{fig:04} in detail, we first consider the Ising model on the Bethe lattices
($\infty,q$). The shortened linear decrease of the squared order parameters
$M^{2}_{(\infty,q)} \propto \left(T_{\rm pt}^{(\infty,q)}-T\right)$ toward the phase
transition points is plotted on the top graph in Fig.~\ref{fig:06} for $q=4,5,6,\dots,13$
and confirms the mean-field nature.

Now we specify the linear divergence of $T_{\rm pt}^{(\infty,q)}\propto q$ in detail.
It can be easily derived in the asymptotic regime for the Ising model on the Bethe
lattice if $q\gg4$ so that
\begin{equation}
T_{\rm pt}^{(\infty,q\gg4)} \to q-1 \, ,
\label{Bethe_asympt}
\end{equation}
where we have made use of
\begin{equation}
T_{\rm pt}^{(\infty,q)} = \frac{1}{\ln\sqrt{\frac{q}{q-2}}}
\equiv \frac{1}{{\rm arctanh}\left(\frac{1}{q-1}\right)} \approx q-1
\end{equation}
if $q\gg4$. The inset of the top graph in Fig.~\ref{fig:06} shows the asymptotic behavior
of $M^{2}_{(\infty,q)}$ with respect to the rescaled horizontal axis $T-(q-1)$. The
data in the inset fully satisfy the limit in Eq.~\eqref{Bethe_asympt}. We support the
data of the Ising model on the lattice geometry ($\infty,100$), which tend to reach the
asymptotic geometry ($\infty,\infty$) as shown in the inset since
$\lim_{q\to\infty}T_{\rm pt}^{(\infty,q)}-(q-1) = 0$.

The numerical data at the phase transition are also verified by the specification
of the linear dependence of the transition temperatures $T_{\rm pt}^{(\infty,q)}$ on $q$.
In general, let us assume a $q$-dependent effective exponent $\alpha_{\rm eff}^{(\infty,q)}$,
\begin{equation}
1 + T_{\rm pt}^{(\infty,q)} \propto q^{\alpha_{\rm eff}^{(\infty,q)}}\, .
\label{alpha}
\end{equation}
The fast convergence of $\alpha_{\rm eff}^{(\infty,q)}\to 1$ with the increasing $q$
is depicted in the inset on the lower right graph in Fig.~\ref{fig:06}, for the
additional data with the coordination numbers $q=20,40,60,80,$ and $100$. The phase
transition temperatures of the Ising model on the Bethe lattices reach the sufficiently
high numerical accuracy with respect to Eq.~\eqref{Bethe_q}. The relative error is
as small as $\varepsilon^{~}_{T_{\rm pt}}\approx10^{-5}$ if calculated at the phase
transition temperature $T_{\rm pt}^{(\infty,q)}$ as shown in the top left inset of
Fig.~\ref{fig:06}. (The inset demonstrates the lowest numerical accuracy, which is
known to occur at phase transitions.)

Up to this point we have verified the correctness of the recurrence relations by
evaluating the phase transition points. We have compared our results with the exact
solutions on the Bethe lattices and evaluated the largest numerical errors at the
phase transitions. In the following, we proceed with the derivation of the
free energy with respect to ($p,q$).

\section{Free energy calculation}

Let the free energy for any lattice geometry ($p,q$) [cf. Eq.~\eqref{part_fnc}]
be normalized per lattice spin site to avoid any divergences associated with the
thermodynamic limit. The free energy per site, expressed as a function of the
iteration step $k$, has the form
\begin{equation}
{\cal F}_{(p,q)}^{[k]} = -\frac{T}{{\cal N}_{(p,q)}^{[k]}}
\ln{\cal Z}_{(p,q)}^{[k]}
\equiv -\frac{T\ln{\rm Tr}\ {\left({\cal C}_k\right)}^q}{{\cal N}_{(p,q)}^{[k]}}\, .
\label{fe}
\end{equation}
The normalization of the free energy per spin site is given by a non-trivial integer
function ${\cal N}_{(p,q)}^{[k]}$, which counts the total number of the spin sites
with respect to given $k$ and ($p,q$). The free energy per site plays a crucial role
in the current analysis since one can derive all the thermodynamic functions from it
in order to determine the phase transition accurately. On the contrary to the
magnetization, the free energy involves the boundary effects.

A direct numerical calculation of the free energy per site in Eq.~\eqref{fe} frequently
results in an extremely fast divergence of the partition function ${\cal F}_{(p,q)}^{[k]}$,
as well as the total number of sites ${\cal N}_{(p,q)}^{[k]}$ on hyperbolic lattices whenever
$k\gtrsim 10$. Therefore, the numerical operations with the tensors ${\cal C}_k$ and
${\cal T}_k$ require to consider an appropriate norm (normalization) in each iteration
step $k$. Two types of the norm are at hand. If introducing a vector ${\cal A}$ with $N$
entries ($a_1 a_2 a_3\dots a_N$), the two norms are
\begin{equation}
\begin{split}
{||{\cal A}||}_1 & =
\max{\left\{ \vert a_i \vert \right\}}_{i=1}^{N}\, ,\\
{||{\cal A}||}_2 & =
\sqrt{\sum\limits_{i=1}^{N} a_i^2}\,.
\end{split}
\label{norms}
\end{equation}
Since any tensor can be "reshaped" into a vector (by grouping all the tensor indices into
the single one), we can also reshape any corner transfer tensor ${\cal C}_1(\sigma_1
\sigma_2\dots\sigma_p)$ with the $p$ tensor indices $\sigma$ and represent the tensor in
the vector form with $N=M^p$ entries, where each index is an $M$-state spin variable
$\sigma=0,1,\dots,M-1$. Any of the two norms is appropriate to choose in the numerical
analysis. We have used the "max" norm ${||\cdot||}_1$ in CTMRG due to its simplicity and
robustness against computational over(under)flows.
Let the positive real numbers $c_k={\vert\vert{\cal C}_k\vert\vert}_1$ and $t_k={\vert
\vert{\cal T}_k\vert\vert}_1$ be the $k$-dependent normalization factors of the respective
tensors ${\cal C}_k$ and ${\cal T}_k$ so that the normalized tensors are written as
\begin{equation}
\begin{split}
{\widetilde{\cal C}}_k & = \frac{{\cal C}_k}{{\vert\vert{\cal C}_k\vert\vert}_1}
                    \equiv \frac{{\cal C}_k}{c_k}\, ,\\
{\widetilde{\cal T}}_k & = \frac{{\cal T}_k}{{\vert\vert{\cal T}_k\vert\vert}_1}
                    \equiv \frac{{\cal T}_k}{t_k}\, .
\end{split}
\end{equation}

For clarity, we again split the free-energy analysis into three parts as we have done above.
First, the free energy per site is derived for an arbitrary $M$-state spin model on the
Euclidean square lattice ($4,4$) by means of the recurrence relations in Eq.~\eqref{rr44}.
In the following, the hyperbolic lattice ($5,4$) is considered, and a recurrence formalism
of the free energy is given for $k=3$, which is associated with a graphical description
of the lattice for the direct visual comparison. Finally, we generalize the free-energy
calculation for any ($p,q$) lattice geometry.

\subsection{Free energy on (4,4) lattice}

\begin{figure}[tb]
{\centering\includegraphics[width=\linewidth]{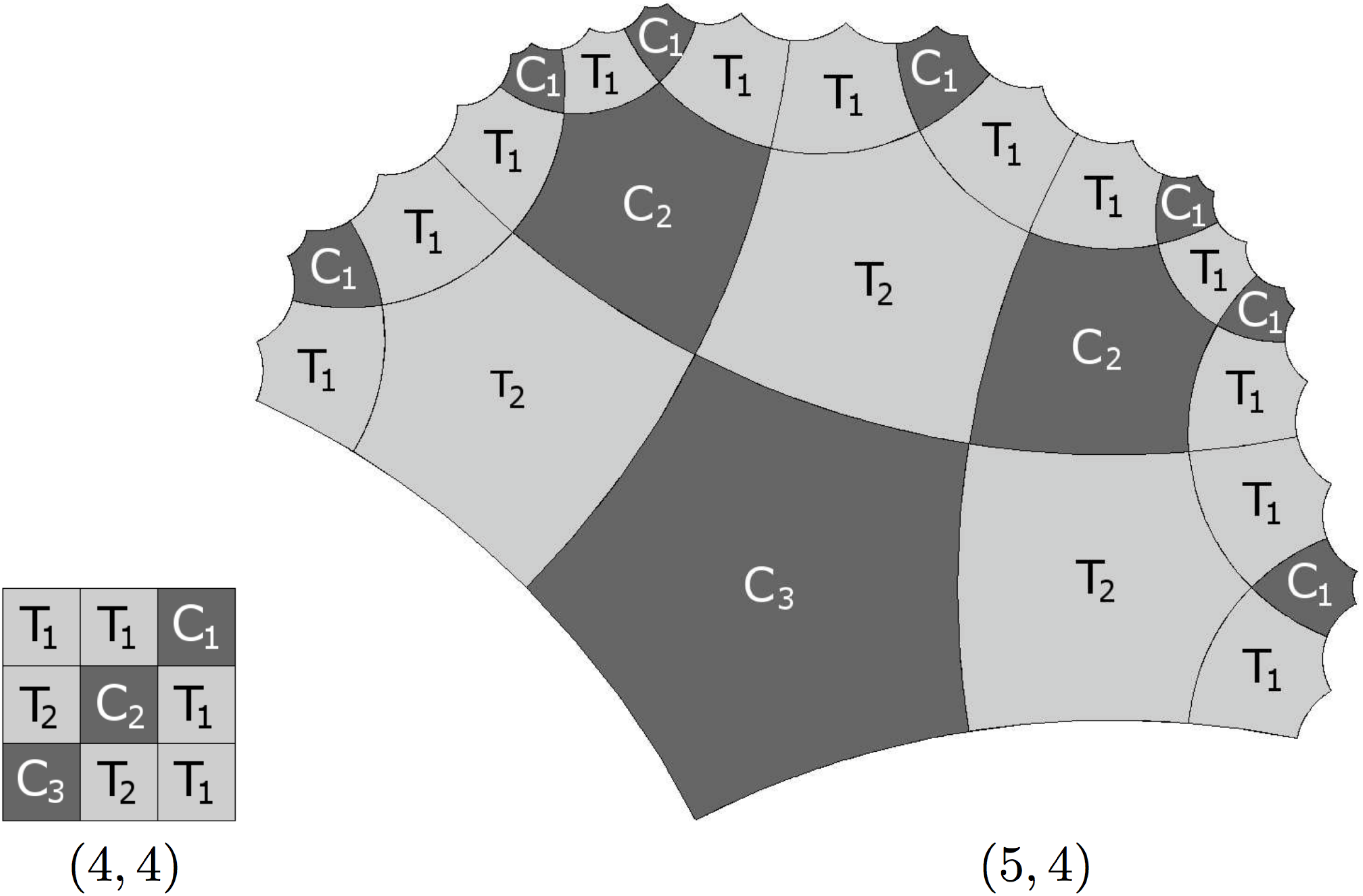}}
  \caption{The graphical representation of the corner transfer tensor ${\cal C}_3$ for
the square lattice ($4,4$) on the left and for the hyperbolic lattice ($5,4$) on the
right, respectively, in accord with Eqs.~\eqref{rr44} and \eqref{rr54} for $k=3$. We
use the dark and the bright shaded $p$-gons in order to distinguish clearly between
${\cal C}_k$ and ${\cal T}_k$, respectively.}
  \label{fig:07}
\end{figure}

The number of the spin sites ${\cal N}_{(4,4)}^{[k]}$ satisfies the power law with respect
to the iteration step $k$ on the square lattice (cf. Fig.~\ref{fig:01}),
\begin{equation}
{\cal N}_{(4,4)}^{[k]} = (2k+1)_{~}^{2}\,.
\label{N44}
\end{equation}
As the lattice size expands its size with growing $k$, the normalization factors start
accumulating recursively. For instance, the explicit expression for the free energy in
the third iteration step gives ${\cal F}_{(4,4)}^{[3]} = -T\ln{\rm Tr}\ ({\cal C}_3)^4/7^2$,
where each of the four central corner tensors ${\cal C}_3$ is recursively decomposed into
the product of the normalized tensors ${\widetilde{\cal C}}_2$ and ${\widetilde{\cal T}}_2$,
which again depend on ${\widetilde{\cal C}}_1$ and ${\widetilde{\cal T}}_1$ with regard to
Eq.~\eqref{rr44}. The recurrence relations result in the nested dependence of the normalization
factors $c_k$ and $t_k$. Hence, the decomposition of one of the four normalized corner
transfer tensors on the square lattice gives
\begin{equation}
\begin{split}
{\widetilde{\cal C}}_3^{~} & = \frac{{\cal C}_3^{~}}{c_3^{~}}
 = \frac{{\cal W}_{\rm B}^{~}{\widetilde{\cal T}}_2^2{\widetilde{\cal C}}_2^{~}}
          {c_3^{~}}
 =\frac{{\cal W}_{\rm B}^{~}{\cal T}_2^2{\cal C}_2^{~}}
             {t_2^2c_2^{~}c_3^{~}} \\
& = \frac{{\cal W}_{\rm B}^{~}{\left({\cal W}_{\rm B}^{~}{\widetilde{\cal T}}_1^{~}\right)}^2
{\left({\cal W}_{\rm B}^{~}{\widetilde{\cal T}}_1^2{\widetilde{\cal C}}_1^{~}\right)}}
{t_2^2 c_2^{~}c_3^{~}}\\
& = \frac{{\cal W}_{\rm B}^4 {\cal T}_1^4{\cal C}_1^{~}}
{t_1^4 t_2^2 c_1^{~}c_2^{~}c_3^{~}}
 = \frac{{\cal W}_{\rm B}^9}
{t_1^4 t_2^2 t_3^0 c_1^{1}c_2^{1}c_3^{1}}\, ,
\end{split}
\label{C_norm_44}
\end{equation}
and it has the identical graphical representation depicted in Fig.~\ref{fig:07} on the left.
Thus, Eq.~\eqref{C_norm_44} can be read in a way where the single corner tensor
${\cal C}_3$ is composed of the nine Boltzmann weights denoted by the power in the
numerator, whereas the denominator counts the multiplicity of the normalization
factors $c_k$ and $t_k$ for $k=1,2,3$, in accord with Fig.~\ref{fig:07}. Substituting
${\cal C}_3$ into Eq.~\eqref{part_fnc}, we get the explicit expression for the free energy
per site when $k=3$
\begin{equation}
{\cal F}_{(4,4)}^{[3]} = -\frac{4T}{7^2}
\ln\left({\rm Tr}\ \widetilde{\cal C}_{3}^{~}\prod\limits_{j=1}^{3}
c_{j}^{~}t_{j}^{2(3-j)}\right)\, .
\end{equation}

The generalized form of the free energy per site for the arbitrary $k$, written in terms
of the normalization factors of the four central tensors ${\widetilde{\cal C}_{k}^{~}}$, is
\begin{equation}
{\cal F}_{(4,4)}^{[k]} = -\frac{4T\ln{\rm Tr}\ \widetilde{\cal C}_{k}^{~}}{(2k+1)_{~}^{2}}
-\frac{4T\sum\limits_{j=0}^{k-1}\ln c_{k-j}^{~} + \ln t_{k-j}^{2j}}{(2k+1)_{~}^{2}}
\label{FE44}
\end{equation}
noticing that the tensors ${\cal C}_k$ and ${\cal T}_k$ functionally depend on the tensors
from the previous iteration steps $k-1,k-2,\dots,1$. To be more specific, for given $k=3$,
the nested dependence, ${\cal C}_3^{~}[{\cal C}_2^{~}({\cal C}_1^{~},{\cal T}_1^{2}),
{\cal T}_2^{2}({\cal T}_1^{2})]$, is explicitly involved in accord with Eq.~\eqref{C_norm_44}.
Hence, the same recurrence nested dependence is inherited into the normalization factors
$c_k$ and $t_k$, which are also functions of temperature.

The normalization we have thus introduced causes that the three terms (real numbers)
$c_k$, $t_k$, and ${\rm Tr}\ \widetilde{\cal C}_{k}^{~}$ converge if $k\to\infty$.
This property is useful for the derivation of a simplified free energy formula with
the known analytic background~\cite{Baxter}. We denote the three converged real values in
the following ${\rm Tr}\ \widetilde{\cal C}_{\infty}^{~} = z$, $c_{\infty} = c$, and
$t_{\infty} = t$. Numerically, it means that from a certain (finite) threshold, the
iteration step $k^{\ast}$, any further increase of the iteration steps above
$k^{\ast}$ does not affect the fixed values so that ${\rm Tr}\,\widetilde{\cal C}_{
k\geq k^{\ast}}^{~}=z$, $c_{k\geq k^{\ast}}^{~}=c$, and $t_{k\geq k^{\ast}}^{~}=t$.
After a short algebra, if considering the thermodynamic limit $k\to\infty$,
the free energy per site in Eq.~\eqref{FE44} is simplified to the final form
\begin{equation}
{\cal F}_{(4,4)}^{[\infty]} = -T\ln t\, .
\end{equation}
Here $t$ is nothing but the largest eigenvalue of the standard transfer matrix~\cite{Baxter}
defined between two adjacent infinite rows (or columns) of spins, which are given by the
infinite product of the four-site (square) Boltzmann weights ${\cal T}_{\infty}^{~}=
\prod{\cal W}_{\rm B}^{~}$. The largest eigenvalue of the standard transfer matrix
${\cal T}_{\infty}^{~}$ is equivalent to the normalization factor $t$ irrespective of
the two norms we have defined in Eq.~\eqref{norms} on the square lattice.

\subsection{Free energy on (5,4) lattice}

We now consider the other example. It is instructive to express graphically the complete
structure of the normalized corner tensor ${\widetilde{\cal C}}_3$ on the hyperbolic
lattice ($5,4$) as depicted in Fig.~\ref{fig:07} on the right. The structure agrees
with the recurrence relations in Eqs.~\eqref{rr54}. The analogous decomposition of
${\widetilde{\cal C}}_3$ into the normalization factors $c_j$ and $t_j$ (for $j=1,2,\dots,k$)
gives
\begin{equation}
\begin{split}
{\widetilde{\cal C}}_3^{~} & = \frac{{\cal C}_3^{~}}{c_3^{~}}
 = \frac{{\cal W}_{\rm B}^{~}{\widetilde{\cal T}}_2^{3}{\widetilde{\cal C}}_2^{2}}
          {c_3^{~}}
 =\frac{{\cal W}_{\rm B}^{~}{\cal T}_2^{3}{\cal C}_2^{2}}
             {t_2^3 c_3^{~}c_2^{2}} \\
& = \frac{{\cal W}_{\rm B}^{~}{\left({\cal W}_{\rm B}^{~}{\widetilde{\cal T}}_1^{2}
{\widetilde{\cal C}}_1^{~}\right)}^3
{\left({\cal W}_{\rm B}^{~}{\widetilde{\cal T}}_1^{3}{\widetilde{\cal C}}_1^{2}\right)}^2}
{t_2^3 c_2^{2} c_3^{~}}\\
& = \frac{{\cal W}_{\rm B}^6{\cal T}_1^{12}{\cal C}_1^{7}}
{t_1^{12} t_2^{3} c_1^{7}c_2^{2}c_3^{~}}
 = \frac{{\cal W}_{\rm B}^{25}}
{t_1^{12} t_2^{3} t_3^0 c_1^{7}c_2^{2}c_3^{1}}\, .
\end{split}
\label{C_norm_54}
\end{equation}
Evidently, the power in the Boltzmann weight completely reproduces the pentagonal lattice
structure ($5,4$) shown in Fig.~\ref{fig:07}. The powers associated with the normalization
factors also coincide with the number of the individual tensors depicted graphically.

We denote the powers in the factors $c_j$ and $t_j$ for $j=1,2,\dots,k$, respectively,
by the integer exponents $n_{k-j+1}$ and $m_{k-j+1}$, which are indexed in the reverse
order for the later convenience of writing the expressions in a simpler form. In
particular, the integer exponents in the denominator of Eq.~\eqref{C_norm_54} satisfy
the ordering $t_1^{m_3} t_2^{m_2} t_3^{m_1} c_1^{n_3}c_2^{n_2}c_3^{n_1}$. They are
also used in the computation of the total number of the spin sites
${\cal N}_{(5,4)}^{[k]}$ via the relation
\begin{equation}
{\cal N}_{(5,4)}^{[k]} = 1 + 4 \sum\limits_{j=1}^{k} 3 n_j + 2 m_j \, .
\label{N54}
\end{equation}
At the same time, the integer exponents $n_j$ and $m_j$ have to satisfy the recurrence
relations
\begin{equation}
\begin{split}
n_{j+1}&=2n_{j}+m_{j}\, , \qquad\ \,\;  n_{1}=1\, , \\
m_{j+1}&=3n_{j}+2m_{j}\, ,\qquad m_{1}=0\, .
\end{split}
\end{equation}
Since the entire lattice ($5,4$) is made by tiling the four corner tensors ${\cal C}_k$
meeting at the central spin site (cf. Fig.~\ref{fig:01}), the number $1$ and the prefactor
$4$ (in front of the summation) in Eq.~\eqref{N54}, respectively, correspond to the central
spin and the four joining tensors ($q=4$). The remaining two prefactors $3$ and $2$ under
the summation in Eq.~\eqref{N54} count those spin sites, which are not shared by the two
attached pentagonal tensors ${\cal C}_j$ and ${\cal T}_j$ ($j=1,2,\dots,k$), respectively.
[A deep analysis of the ($5,4$) lattice is inevitable to understand all the details.]

\begin{figure}[tb]
{\centering\includegraphics[width=\linewidth]{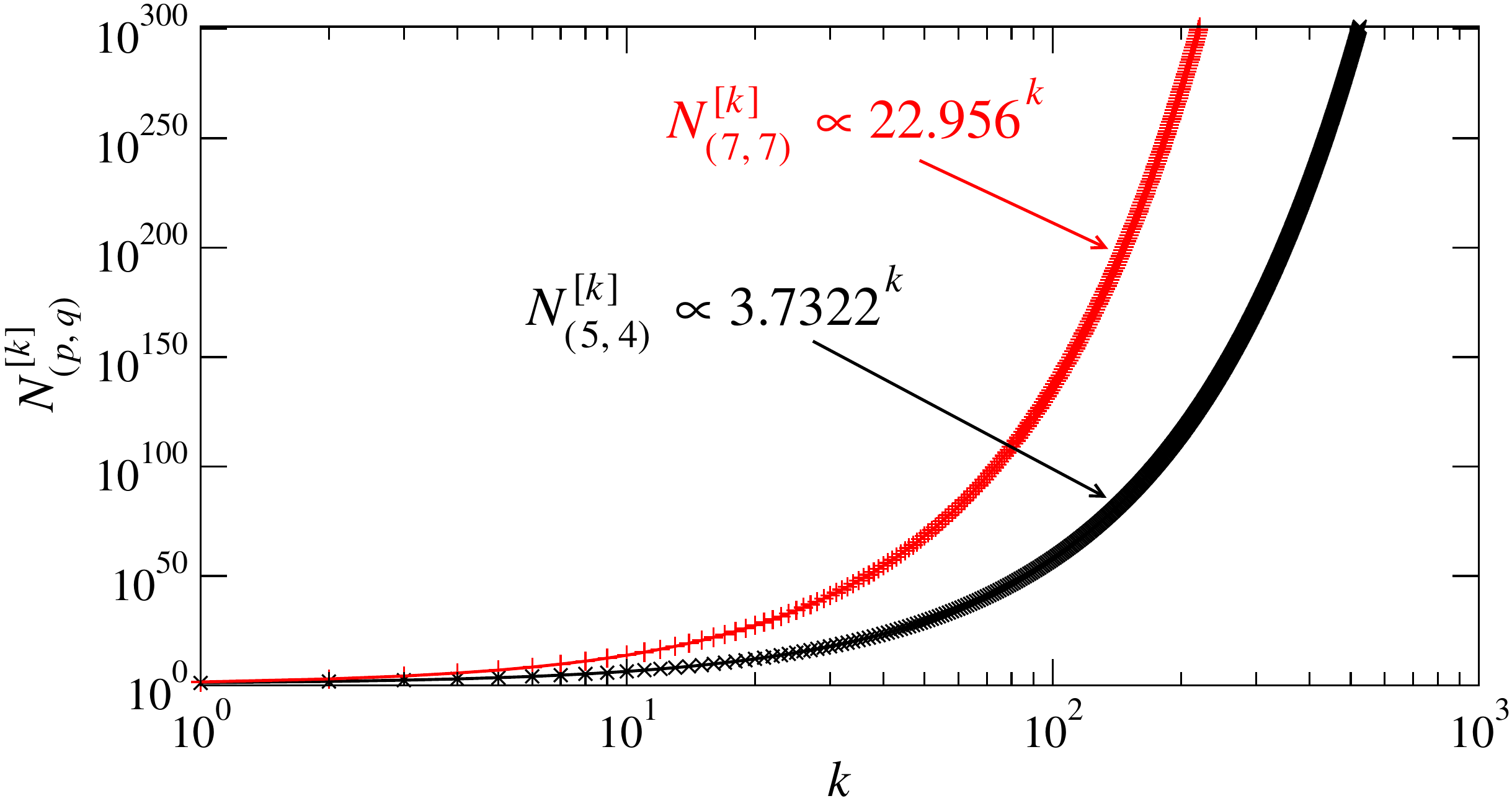}}
  \caption{(Color online) The exponential dependence of the total number of the spins
on the iteration number $k$ (the least-squares fitting in the log-log plot) for the two
lattices ($5,4$) and ($7,7$).}
\label{fig:08}
\end{figure}

The free energy per site at given $k$ on the hyperbolic lattice ($5,4$) can be expressed
in the following generalized form:
\begin{equation}
{\cal F}_{(5,4)}^{[k]} =- \frac{4T\ln{\rm Tr}\ \widetilde{C}_{k}^{~}}
      {{\cal N}_{(5,4)}^{[k]}}
-\frac{4T\sum\limits_{j=0}^{k-1} \ln c_{k-j}^{n_{j+1}} + \ln t_{k-j}^{m_{j+1}}}
      {{\cal N}_{(5,4)}^{[k]}}\, .
\label{FE54}
\end{equation}
The first term converges to zero with increasing $k$, i.e.,
\begin{equation}
\lim\limits_{k\to\infty} \frac{4T\ln{\rm Tr}\ \widetilde{C}_{k}^{~}}
                              {{\cal N}_{(5,4)}^{[k]}}  =  0 \, ,
\end{equation}
since the normalized partition function $\widetilde{\cal Z}_{(p,q)}^{[\infty]}
\equiv {\rm Tr}\ \widetilde{C}_{\infty}^{~}$ is bounded at any temperature in
the thermodynamic limit,
\begin{equation}
1 \leq \widetilde{\cal Z}_{(p,q)}^{[\infty]} \leq M\, ,
\label{FElim}
\end{equation}
for an arbitrary $M$-state spin system. The lower and the upper bounds correspond
to the limits $\lim_{T\to 0}\widetilde{\cal Z}_{(p,q)}^{[\infty]} = 1$ and
$\lim_{T\to\infty}\widetilde{\cal Z}_{(p,q)}^{[\infty]} = M$, respectively.
Finally, the number of spin sites in the denominator of the first term grows
exponentially ${\cal N}_{(5,4)}^{[k]} \propto 3.7^k$, as plotted in Fig.~\ref{fig:08}.

\subsection{Free energy on ({\boldmath{$p,q$}}) lattices}

The generalization of the free-energy calculation for any (multi-state) spin model
on an arbitrary lattice geometry ($p\geq4,q\geq4$) is straightforward and requires
a careful graphical analysis of many lattice geometries, which is beyond the scope
of this work for its extensiveness. The free energy per spin for a finite $k$ has
the generalized form
\begin{equation}
\begin{split}
{\cal F}_{(p,q)}^{[k]} & = -\frac{qT\ln{\rm Tr}\ \widetilde{C}_{k}^{~}}
      {{\cal N}_{(p,q)}^{[k]}}
-\frac{qT\sum\limits_{j=0}^{k-1} \ln c_{k-j}^{n_{j+1}} + \ln t_{k-j}^{m_{j+1}}}
      {{\cal N}_{(p,q)}^{[k]}} \\
& \overset{k\gg1}{=} -\frac{qT\sum\limits_{j=0}^{k-1}n_{j+1} \ln c_{k-j}^{~}+m_{j+1}\ln t_{k-j}^{~}}
      {{\cal N}_{(p,q)}^{[k]}}\, ,
\end{split}
\label{FEpq}
\end{equation}
where the total number of the spin sites is expressed as
\begin{equation}
{\cal N}_{(p,q)}^{[k]} = 1 + q \sum\limits_{j=1}^{k} (p-2) n_j + (p-3) m_j\, ,
\label{Npq}
\end{equation}
and the integer variables $n_j$ and $m_j$ satisfy more complex recurrence relations,
\begin{equation}
\begin{split}
n_{j+1} = &\, [(p-2)(q-3)-1]n_{j} + [(p-3)(q-3)-1]m_{j}\, ,\\
m_{j+1} = &\, (p-2)n_{j}+(p-3)m_{j}\, ,\\
 n_{1} = & \, 1\, ,\\
 m_{1} = & \ 0.
\end{split}
\label{nmpq}
\end{equation}
The evaluation of Eq.~\eqref{Npq} is carried out numerically, and a strong exponential
behavior occurs with increasing $p$ and $q$. Figure~\ref{fig:08} shows the log-log plot
of the exponential increase of the total number of the sites ${\cal N}_{(7,7)}^{[k]}
>{\cal N}_{(5,4)}^{[k]}$.

The final expression for the free energy in Eqs.~\eqref{FEpq}-\eqref{nmpq} also includes
the case of the Euclidean lattice ($4,4$). The complete equivalence with Eq.~\eqref{FE44}
can be easily verified if considering that $n_j=n_{j-1}=\cdots=n_1\equiv1$
and $m_j=2n_{j-1}+m_{j-1}=2(j-1)+m_1\equiv 2j-2$, which reduce the exponential dependence
of the total number of the spin sites back to the power-law behavior in Eq.~\eqref{N44},
\begin{equation}
{\cal N}_{(4,4)}^{[k]} = 1 + 4 \sum\limits_{j=1}^{k} 2 n_j + m_j \equiv {(2k+1)}^2_{~}\, .
\end{equation}

\section{Results}

Having defined the free energy per site for the lattice geometries ($p,q$), we analyze
the phase transition of spin models in the thermodynamic limit on the four representative
lattices ($4,4$), ($4,7$), ($7,4$), and ($7,7$) we have used earlier. We have shown that
the phase transition temperatures $T_{\rm pt}$, calculated by the spontaneous magnetization
$M_{(p,q)}$ at the lattice center, correctly reflected the bulk properties, and the
boundary effects were eliminated. In other words, if various types of the boundary
conditions (such as free and fixed ones) are imposed, the phase transition of the Ising
model is not affected provided that we evaluated the expectation value $\langle
\sigma_c\rangle$ in Eq.~\eqref{Mpq}. The correctness and high numerical accuracy has
been compared with the exactly solvable Ising model on Bethe lattice~\cite{Baxter}.

\begin{figure}[tb]
{\centering\includegraphics[width=\linewidth]{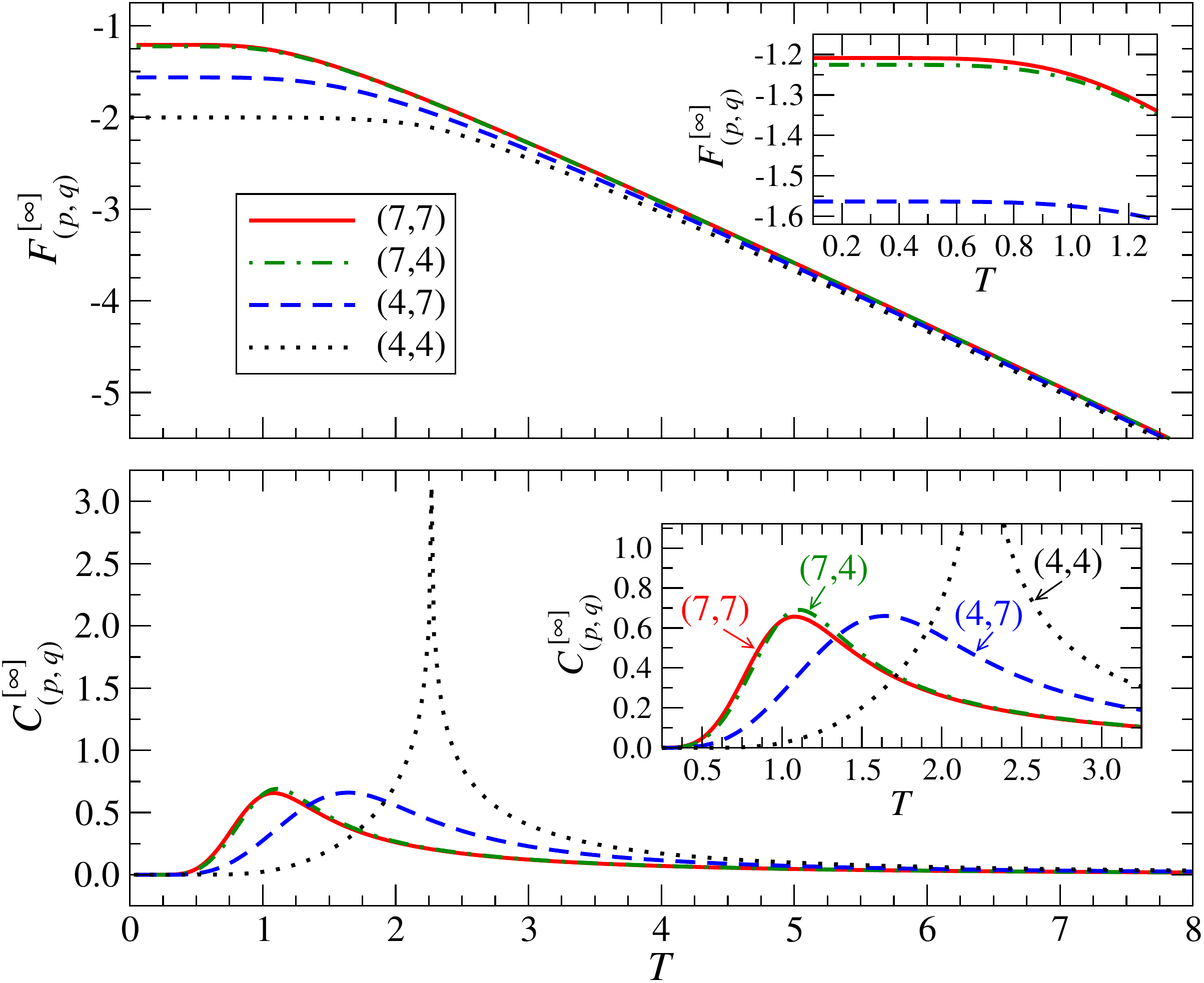}}
  \caption{(Color online) The free energy from Eq.~\eqref{FEpq} and the specific heat
from Eq.~\eqref{Cpq} vs temperature on the selected lattices ($4,4$), ($4,7$), ($7,4$),
and ($7,7$). The inset of the bottom graph shows the details of the broadened specific
heat maxima.}
\label{fig:09}
\end{figure}

\subsection{Absence of phase transition on non-Euclidean lattices}

The top graph of Fig.~\ref{fig:09} shows the free energies on the four representative
lattices, which are evaluated by Eq.~\eqref{FEpq}. A phase transition can be associated with
a singular (non-analytic) behavior of the specific heat being the second derivative of
the free energy with respect to temperature,
\begin{equation}
C_{(p,q)}^{[\infty]} = -T\frac{\partial^2}{\partial T^2}
{\cal F}_{(p,q)}^{[\infty]}\, .
\label{Cpq}
\end{equation}
The temperature dependence of the specific heat of the Ising model on the respective
four lattices is plotted on the bottom graph in Fig.~\ref{fig:09}. Evidently, we find
the non-analytic behavior on the square lattice ($4,4$) with the diverging peak
at the temperature, which corresponds to the exact critical temperature $T_{\rm c} =
2/\ln(1+\sqrt{2})$.

However, none of the three hyperbolic lattice geometries results in an analogous
non-analytic peak at the phase transition temperatures $T_{\rm pt}^{(p,q)}$
we had calculated from the spontaneous magnetization plotted in Fig.~\ref{fig:03}.
Instead, a broad maximum appears for the particular lattices, which does not correspond
to the correct phase transition temperatures we had detected earlier.

Strong boundary effects on the hyperbolic lattices prevent the Monte Carlo (MC)
simulations from the accurate analysis of phase transition phenomena on the hyperbolic
lattices~\cite{MC1,MC2,MC3,MC4}. The necessity to subtract a couple of boundary site
layers were performed to detect the correct bulk properties~\cite{boundary}.
If defining a ratio of the boundary sites to the total number of sites, the ratio
converges to zero in the Euclidean case, whereas it goes to non-zero values on the
hyperbolic lattices in the thermodynamic limit.

Notice that we had previously studied the phase transition phenomena by means of the
order parameter (the spontaneous magnetization), the nearest-neighbor correlation
function (the internal energy), and the entanglement entropy. These thermodynamic
functions were evaluated at the central part of the infinitely large hyperbolic
lattices, where the effects of the boundary conditions were explicitly disregarded.

\subsection{The Bulk Free Energy}

In order to specify the phase transition temperature on the hyperbolic lattices
correctly, the free energy has to be modified by reducing the boundary layers from
the total free energy. With each next iteration step $k+1$, the CTMRG algorithm expands
the lattice size by increasing and pushing the boundary sites farther from the lattice
center. This expansion process can be regarded as an additional spin layer (the shell),
which is composed of the tensors ${\cal C}_{1}$ and ${\cal T}_{1}$ multiplying its number
in accord with the recurrence relations, and the $q$ tensors ${\cal C}_{k+1}$ are included
in the center of the lattice at the same time; cf. Fig.~\ref{fig:07}. Hence, the lattice
can be thought of as a concentric system of the shells indexed by $j$ so that the
$j^{\rm th}$ shell contains the spin sites, which separate the spin sites in the
tensors ${\cal C}_j$ and ${\cal T}_j$ from the spin sites in the tensors ${\cal C}_{j-1}$
and ${\cal T}_{j-1}$ on a given ($p,q$) geometry (cf. Fig.~\ref{fig:07}).
Such a structure of the concentric shells in the $k^{\rm th}$ iteration step ascribes
the outermost shell to $j=1$ toward the innermost (non-trivial) shell $j=k$ (leaving
the central spin site apart), which is related to the way of counting the total number 
of the spin sites in Eq.~\eqref{Npq}.

If the integer $\ell$ denotes the $\ell$ outermost shells $j=1,2,\dots,\ell<k$, we can
introduce a new quantity, the {\it bulk\,} free energy ${\cal B}_{(p,q)}^{[k,\ell]}$,
which defines the free energy of the $k-\ell$ inner shells. It is given by the subtraction
of the free energy contributing from the $\ell$ outer shells from the total free energy.
In particular, the bulk free energy in the $k^{\rm th}$ iteration step is
\begin{equation}
{\cal B}_{(p,q)}^{[k,\ell]} = {\cal F}_{(p,q)}^{[k]} - {\cal F}_{(p,q)}^{\ast\,[k,\ell]}\, ,
\end{equation}
where the asterisk in the second term denotes the free energy of the $\ell$ outermost
shells so that
\begin{equation}
{\cal F}_{(p,q)}^{\ast\,[k,\ell]} = 
-\frac{qT\sum\limits_{j=k-\ell}^{k-1}n_{j+1} \ln c_{k-j}^{~}+m_{j+1}\ln t_{k-j}^{~}}
      {{\cal N}_{(p,q)}^{\ast\,[k,\ell]}}
\label{FEpq_b}
\end{equation}
and
\begin{equation}
{\cal N}_{(p,q)}^{\ast\,[k,\ell]} = q \sum\limits_{j=k-\ell+1}^{k}
(p-2) n_j + (p-3) m_j\, .
\label{Npq_b}
\end{equation}
For the tutorial purpose, we set $\ell=\frac{k}{2}$ and study the thermodynamic limit
$k\to\infty$. (The non-trivial dependence of the bulk free energy on $\ell$ is to be
thoroughly studied elsewhere~\cite{Yoju}.) Following the remarks below Eq.~\eqref{FEpq}
and without loss of generality, we omit the first term in Eq.~\eqref{FEpq_b} as the term
converges to zero after a few iterations.

\begin{figure}[tb]
{\centering\includegraphics[width=\linewidth]{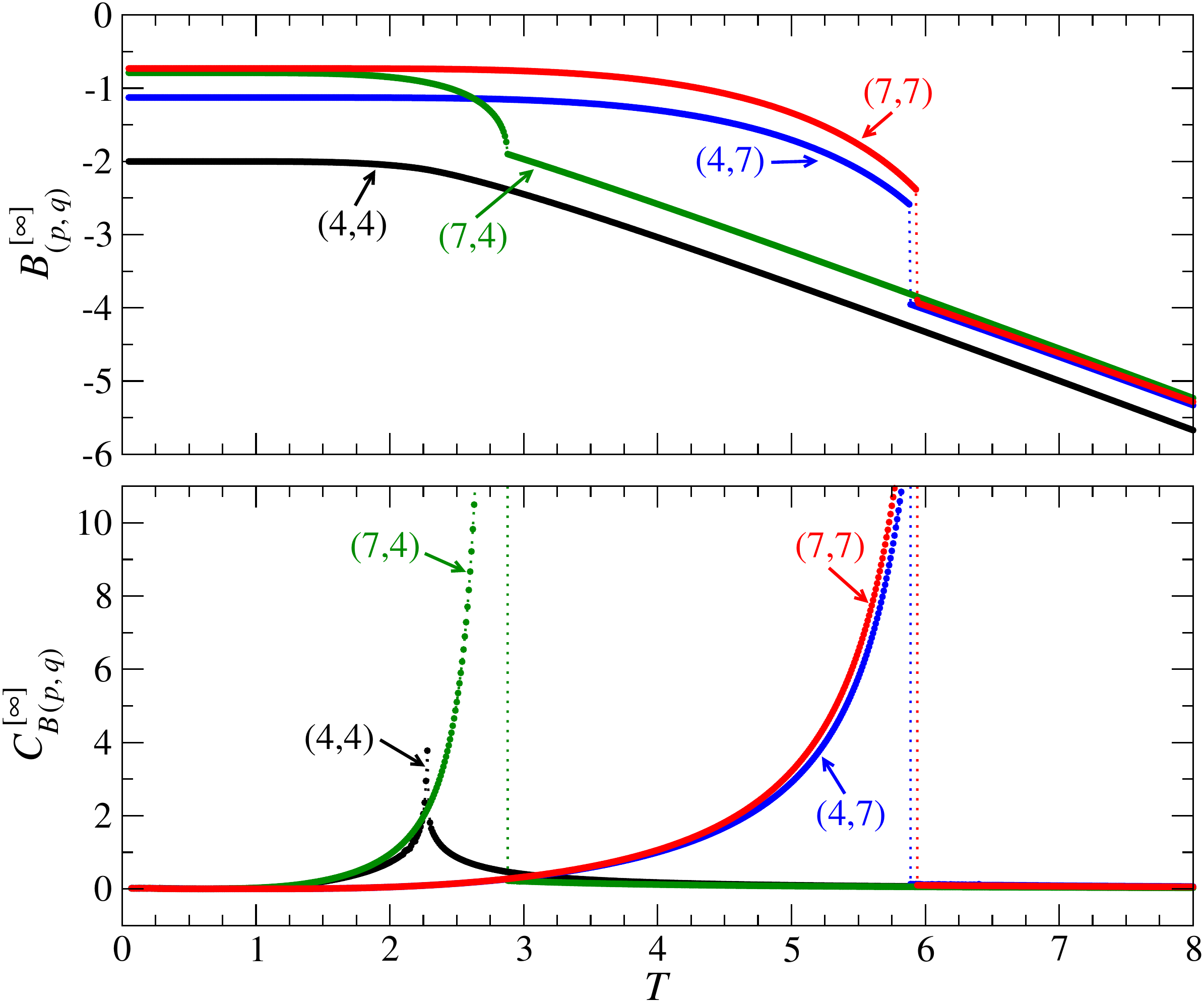}}
  \caption{(Color online) The bulk free energy (the top graph) and the bulk specific
heat (the bottom graph) vs. temperature for the Ising model on the lattices ($4,4$),
($4,7$), ($7,4$), and ($7,7$). The vertical dotted lines accurately correspond to the
phase transition temperatures we have obtained from the spontaneous magnetization in
Fig.~\ref{fig:03}.}
\label{fig:10}
\end{figure}

Figure~\ref{fig:10} (the top graph) shows the bulk free energy for the Ising model
on the four representative lattices in the thermodynamic limit, i.e.,
\begin{equation}
{\cal B}_{(p,q)}^{[\infty]} \equiv
\lim\limits_{k\to\infty} {\cal B}_{(p,q)}^{[k,k/2]}\,.
\end{equation}
In the case of the Euclidean lattice, we get ${\cal B}_{(4,4)}^{[\infty]} \equiv
{\cal F}_{(4,4)}^{[\infty]}$ since the thermodynamic properties in the bulk are not
affected by the boundary conditions. However, the bulk free energy calculated on the
hyperbolic lattices exhibits a remarkable singularity occurring exactly at the phase
transition temperature. The typical structure of the free energy does not change
irrespective of the type of the boundary conditions applied (free and fixed ones).
The maxima of the bulk specific heat,
\begin{equation}
C_{B\,(p,q)}^{[\infty]} = -T\frac{\partial^2}{\partial T^2}
{\cal B}_{(p,q)}^{[\infty]}\, ,
\end{equation}
plotted in the bottom graph of Fig.~\ref{fig:10} accurately correspond to the phase
transition temperatures $T_{\rm pt}^{(p,q)}$ we have studied in Sec.~II. The discontinuous
jump of the bulk specific heat is associated with the typical mean-field universality
behavior~\cite{hCTMRG3q,hCTMRG3qn}, and the vertical dotted lines serve as guides
for the eye to help locate the phase transition temperature. It is worth pointing
out the identical determination of the phase transition temperatures, as we have
obtained independently by the spontaneous magnetization in Fig.~\ref{fig:03}. 

Our definition of the bulk free energy contains lots of interesting features.
For instance, the $\ell$-dependence enables us to observe
and explain the way how the lattice boundary affects the central bulk part of the
lattice and the phase transition if an additional magnetic field is imposed on the
boundary spins only. Such a study is to be published elsewhere~\cite{Yoju}. We do
not follow the Baxter's proposal of calculating the free energy, which is defined
by numerical integrating the spontaneous magnetization with respect to a magnetic
field~\cite{Baxter}. Although such an approach can be numerically feasible and can
be also used to reproduce all the well-known results for the Bethe lattices, it
cannot reflect many aspects of the boundary effects which play the significant role
on the hyperbolic ($p,q$) geometries.

\subsection{Free energy versus lattice geometry}

Having been motivated by the correspondence between the anti-de Sitter spaces and
the conformal field theory of the quantum gravity physics, one can put the question:
{\it "Given an arbitrary spin system on an infinite set of ($p,q$) geometries,
which lattice geometry minimizes the free (ground-state) energy?"}. This is
certainly a highly non-trivial task to be explained thoroughly. Nevertheless, we
attempt to answer the question in the following for a particular set of curved
lattice surfaces we have been considering. This helps us give an insight into the
role of the space geometry with respect to the microscopic description of the
spin interacting system. Although we currently consider the free energy of the
{\it classical} spin lattice systems, we have been recently studying the ground-state
energy of the {\it quantum} spin systems on the lattices ($p\geq4,4$), which also
exhibit qualitatively identical features as studied in this work~\cite{TPVF54,TPVFp4}.
For this reason, the free energy for classical spin systems and the ground-state energy
of quantum spin systems are mutually related.

The free energy per site ${\cal F}_{(p,q)}^{[\infty]}$ converges to a negative
value ${\cal F}_{(p,q)}^{[\infty]}<0$ at finite temperatures $T<\infty$ in the
thermodynamic limit. Scanning the entire set of the ($p\geq4,q\geq4$) geometries,
we show in the following that the free energy per site reaches its minimum on the
square lattice only
\begin{equation}
{\cal F}_{(4,4)}^{[\infty]} = \min_{(p\geq4,q\geq4)}
\left\{{\cal F}_{(p,q)}^{[\infty]}\right\}
\end{equation}
at any fixed temperature $T$. Therefore, we plot the {\it shifted} free energy per
site, ${\cal F}_{(p,q)}^{[\infty]}-{\cal F}_{(4,4)}^{[\infty]}\geq0$, for clarity of
the figures.

\begin{figure}[tb]
{\centering\includegraphics[width=\linewidth]{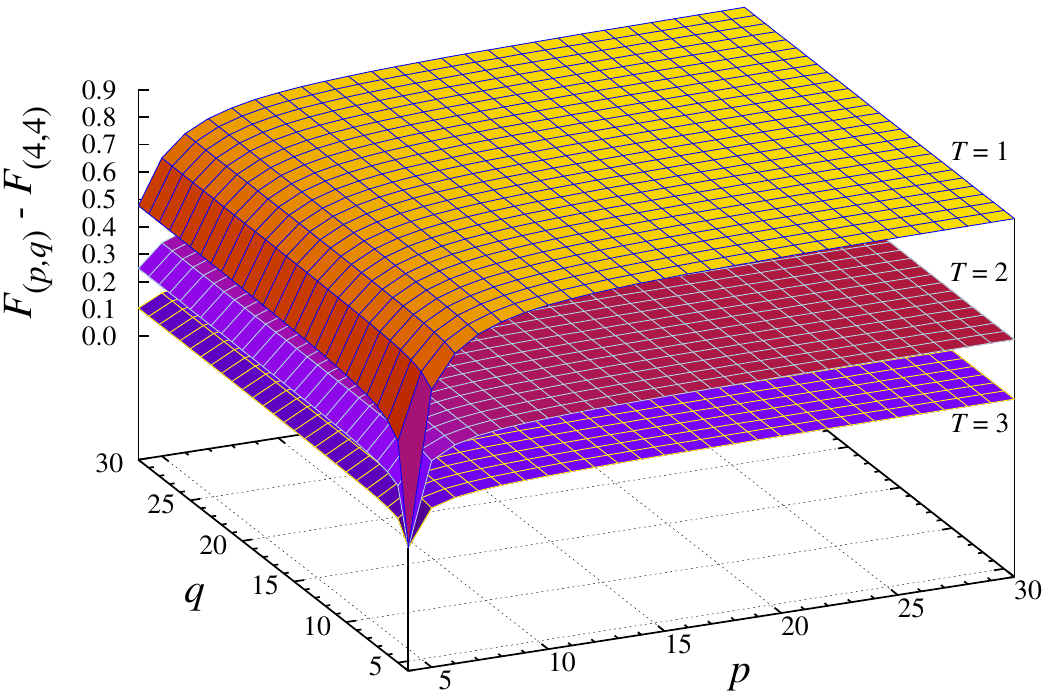}}
  \caption{(Color online) The free energy per site as a function of the lattice geometry
($p,q$) at the selected lower temperatures $T=1,2$, and $3$.}
\label{fig:11}
\end{figure}

\begin{figure}[!b]
{\centering\includegraphics[width=\linewidth]{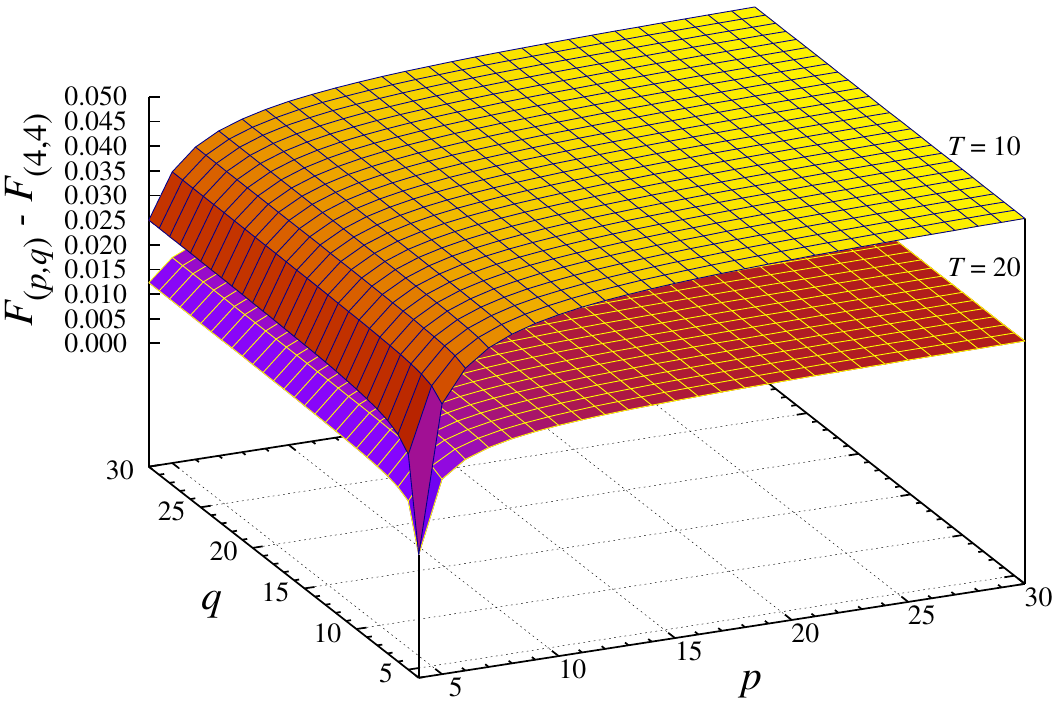}}
  \caption{(Color online) The same as in Fig.~\ref{fig:11} for $T=10$ and $20$.}
\label{fig:12}
\end{figure}

Figures~\ref{fig:11} and \ref{fig:12} show the shifted free energy for the Ising ($M=2$)
model at lower and higher temperatures, respectively. These numerical calculations
unambiguously identify the square lattice geometry, which minimizes the free energy
per spin site. The free energy per site at fixed $T$ becomes less sensitive for higher
values of $p$ and $q$. We observe a weak increasing tendency in the free energy if $p$
increases at fixed $q$; it grows logarithmically as discussed later. The free energy
gets saturated to a constant in the opposite case when $q$ increases at fixed $p$. It is
worth mentioning that the presence of the phase transition does not affect the free
energy minimum observed on the Euclidean square lattice. Moreover, as the temperature
grows, the difference between the free energies on the square and the hyperbolic lattices
weakens.

\begin{figure}[tb]
{\centering
\includegraphics[width=\linewidth]{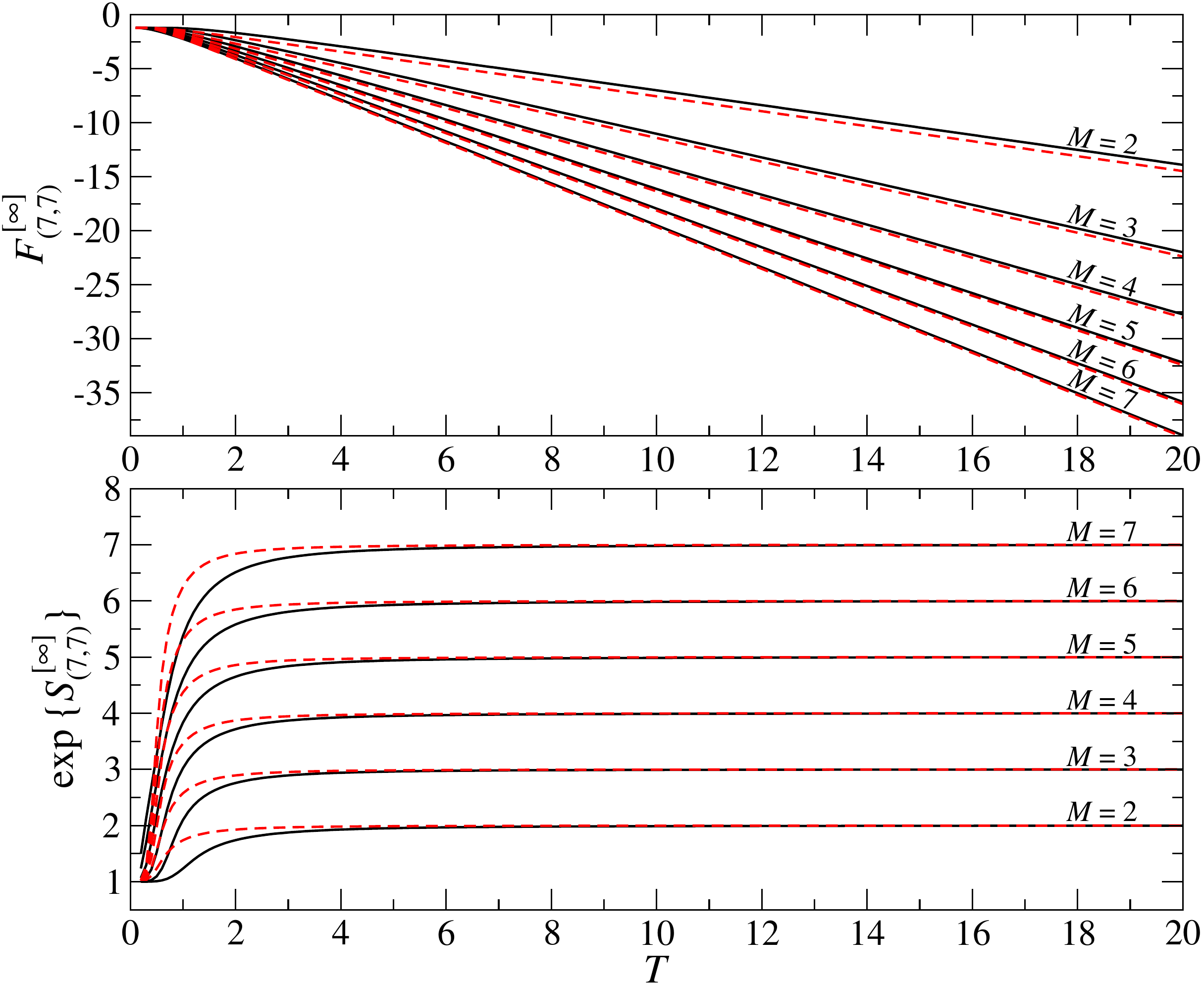}}
  \caption{(Color online) The high-temperature asymptotics of free energy per site
(the top graph) and the entropy (the bottom graph) applied to the lattice ($7,7$).
The full and the dashed lines correspond to the $M$-state clock and the $M$-state
Potts models, respectively, where $M=2,3,\dots,7$.}
  \label{fig:13}
\end{figure}

It is instructive to generalize the free energy features of the Ising model to the other
spin models, the $M$-state clock and $M$-state Potts models. The high-temperature
asymptotics of the free energy for the multi-state spin models on the referenced
hyperbolic lattice ($7,7$) is depicted on the top graph of Fig.~\ref{fig:13}.
The free energy exhibits an asymptotic behavior at higher temperatures for a fixed
lattice geometry ($p,q$). It is a consequence of the thermodynamic limit measured
deeply in the disordered phase, where $T\gtrsim T_{\rm pt}^{(p,q)}\approx q$. Then,
the tensors ${\cal C}_{\infty}$ and ${\cal T}_{\infty}$ prefer higher symmetries
(on the contrary, fewer symmetries occur in the ordered phase if the spontaneous
symmetry breaking is present). The higher symmetries cause that the normalization
factors $c_{k\to\infty} \to M^{p-2}$ and $t_{k\to\infty} \to M^{p-3}$ above $T\gtrsim 2q$,
recalling that the exponents $p-2$ and $p-3$ are associated with the number of the summed
up $M$-state spins in the tensors [cf. also Eq.~\eqref{Npq}]. Substituting $c_{k} =
M^{p-2}$ and $t_{k} = M^{p-3}$ into Eq.~\eqref{FEpq}, one obtains the high-temperature
expression
\begin{equation}
\lim\limits_{{k\to\infty}\atop{T\gtrsim 2q}} {\cal F}_{(p,q)}^{[k]} \propto -T \ln M\, .
\label{F_inf_T}
\end{equation}
The asymptotic linearity of the free energy is examined by the entropy
\begin{equation}
{\cal S}_{(p,q)}^{[\infty]} = -\frac{\partial {\cal F}_{(p,q)}^{[\infty]}}{\partial T}\, ,
\label{Spq}
\end{equation}
which gets saturated so that ${\cal S}_{(p,q)}^{[\infty]} \to \ln (M)$ at $T\gtrsim 2q$.
The lower graph in Figure~\ref{fig:13} satisfies this condition and can be considered
as an independent confirmation of the correct calculation of the free energy per site
in Eq.~\eqref{FEpq}. The high-temperature asymptotic behavior of the entropy is
explicitly plotted to show that $\exp\left\{{\cal S}_{(p,q)}^{[\infty]}\right\}=M$ for
the given set of the $M$-state spin models.

\begin{figure}[tb]
{\centering
\includegraphics[width=\linewidth]{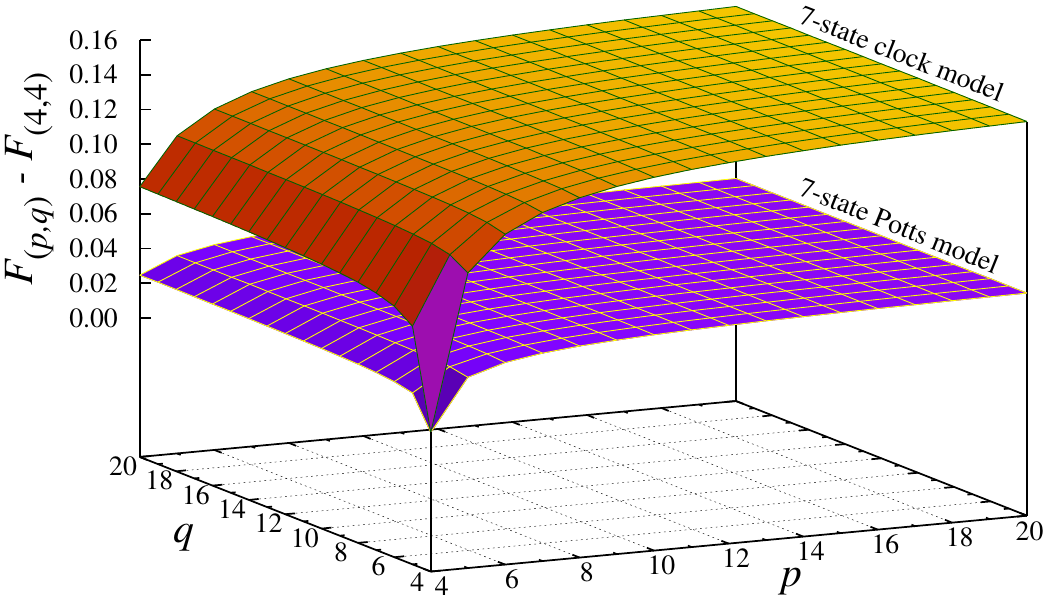}}
  \caption{(Color online) The free energy for the $7$-state clock and Potts models at
$T=5$, which do not differ from the Ising model ($M=2$) in Figs.~\ref{fig:11} and
\ref{fig:12} qualitatively.}
  \label{fig:14}
\end{figure}

Figure~\ref{fig:14} shows the free energy per site with respect to ($p,q$) for the
$7$-state clock and the $7$-state Potts models at $T=5$. Clearly, the free energy
reaches its minimum on the square lattice for the both spin models. Having scanned
the multi-state spin variables $M=2,3,\dots,7$ (not shown) for various temperatures
$T$, the free energy remains minimal on the square lattice ($4,4$).

\subsection{Relation between energy and curvature}

The studied ($p,q$) lattices can be exactly characterized by the radius of Gaussian
curvature~\cite{Mosseri}, which has the analytical expression
\begin{equation}
 {\cal R}_{(p,q)} = -\frac{1}{2\,{\rm arccosh}
            \left[ 
                  \frac{\cos \left( \frac{\pi}{p} \right)}
                       {\sin \left( \frac{\pi}{q} \right)}
            \right]} .
\label{Rpq}
\end{equation}
For later convenience we include the negative sign in ${\cal R}_{(p,q)}$. The radius
of curvature for the square lattice geometry ($4,4$) diverges, $R_{(4,4)}\to-\infty$,
while the remaining hyperbolic lattice geometries ($p,q$) are finite and non-positive.
The analytical description in Eq.~\eqref{Rpq} results in a constant and position
independent curvature at any position on the infinitely large lattices ($p,q$).
It is a consequence of the constant distance between the lattice vertices for all
geometries ($p,q$), which is equivalent to keeping the spin-spin coupling to be $J=1$
in all the numerical analysis of the spin systems on the ($p,q$) lattices.

\begin{figure}[tb]
{\centering\includegraphics[width=\linewidth]{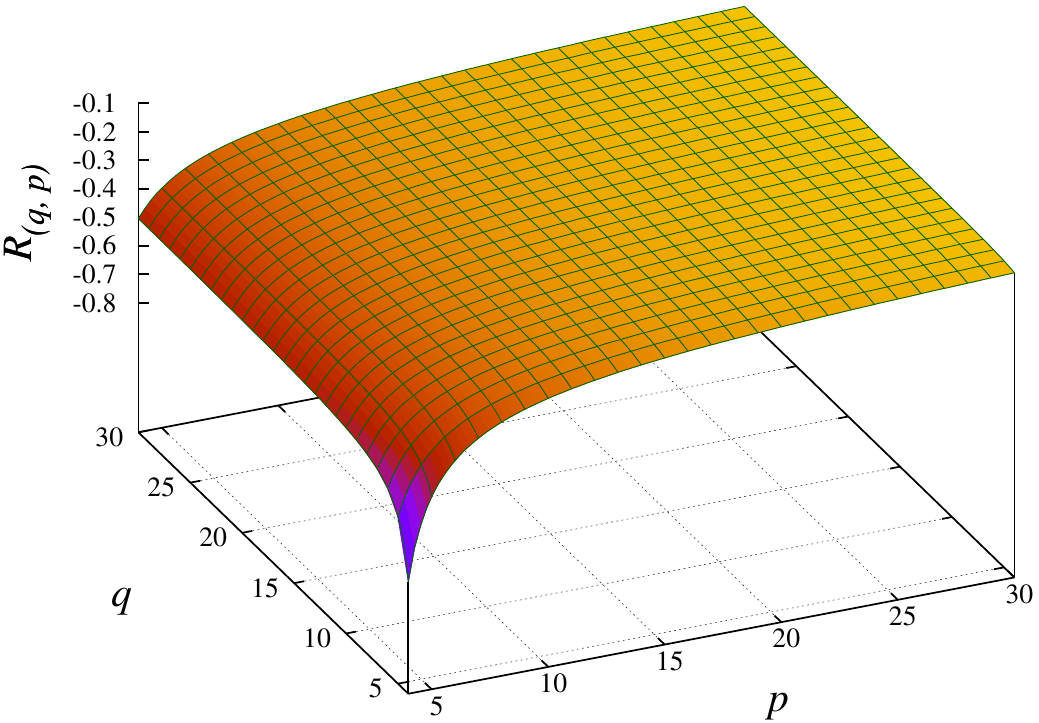}}
  \caption{(Color online) The functional dependence of the Gaussian radius of
curvature ${\cal R}_{(q,p)}$ plotted in the dual lattice geometry ($q,p$).}
\label{fig:15}
\end{figure}

In Fig.~\ref{fig:15} we plot the radius of curvature in the dual geometry ($q,p$),
i.e., the roles of $p$ and $q$ are swapped. It is immediately evident that the surface
shape of ${\cal R}_{(q,p)}$ exhibits a qualitative similarity if compared to the free
energy per site ${\cal F}_{(p,q)}^{[\infty]}$ we depicted in Figs.~\ref{fig:11},
\ref{fig:12}, and \ref{fig:14}.

Such a surprising observation opens new questions about the relation between the energy
at thermal equilibrium and the space (lattice) geometry, which is equivalent to the
relation between the ground-state energy of quantum systems and the underlying geometry.
We, therefore, focus on the low-temperature regime, $T<1$, where the similarity is most
striking, provided that the numerical computations remain reliable in order to avoid any
under- or overflows in the transfer tensors. For this reason, the numerical calculations
require the setting of the 34 decimal digit precision.

\begin{figure}[!b]
{\centering\includegraphics[width=\linewidth]{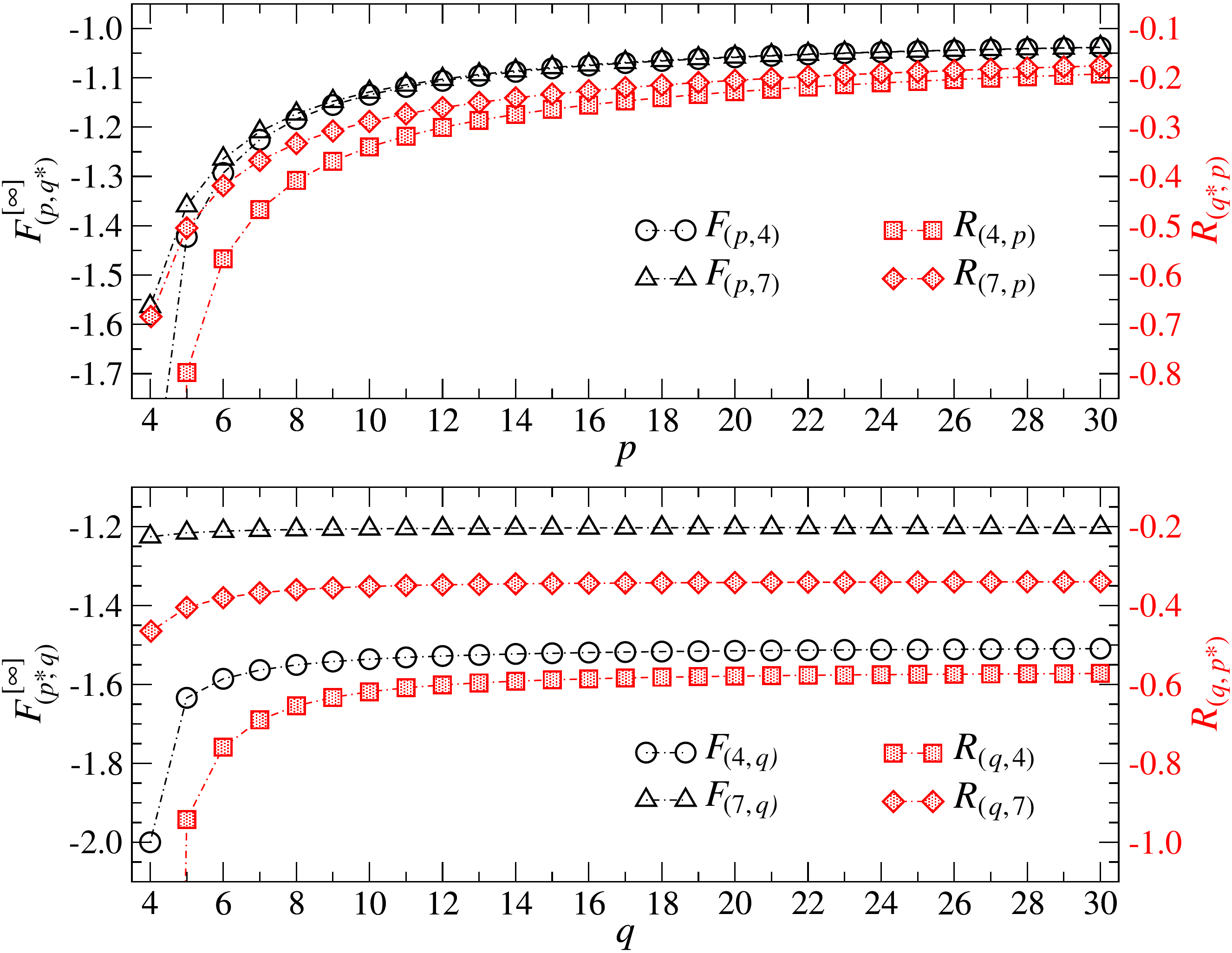}}
  \caption{(Color online) The comparison of the free energy per site for the
Ising model at $T=0.5$ with the Gaussian radius of curvature on the dual geometry.
The top graph shows the case of the two fixed coordination numbers $q^{\ast}=4$
and $q^{\ast}=7$, whereas the lower graph depicts the opposite case when fixing the
$p$-gons to the sizes $p^{\ast}=4$ and $p^{\ast}=7$.}
\label{fig:16}
\end{figure}

In Fig.~\ref{fig:16} we plot both the free energy per site ${\cal F}_{(p,q)}^{[\infty]}$
at $T=0.5$ and the radius of curvature ${\cal R}_{(q,p)}$ on the dual lattice with respect
to $p$ (the top graph) and $q$ (the lower one), while the other associated lattice parameter
is fixed. The top graph compares the free energy per site with the radius of curvature
at fixed $q^{\ast}=4$ and $q^{\ast}=7$ when $4\leq p \leq 30$. On the other hand, the
lower graph displays ${\cal F}_{(p^{\ast},q)}$ and ${\cal R}_{(q,p^{\ast})}$ at fixed
$p^{\ast}=4$ and $p^{\ast}=7$ for $4\leq q \leq 30$. While in the former case both the
functions increase with $p$, in the latter case the functions saturate and converge to
constants.

It is instructive to inspect the asymptotic behavior of ${\cal R}_{(q,p)}$. If $q$ is
fixed to an arbitrary $q^{\ast}\geq4$, the logarithmic dependence on $p$ is present and
${\cal R}_{(q^{\ast},p\gg4)}\to -1/2\ln[\frac{2p}{\pi}\cos(\frac{\pi}{q^{\ast}})]$.
Fixing $p$ to $p^{\ast}$ causes that the radius of curvature converges to a constant
so that for a sufficiently large $p^{\ast}$, the constant does not depend on $q$ and
${\cal R}_{(q\gg4,p^{\ast})} \to -1/2\ln[\frac{2}{\sin(\pi/p^{\ast})}
-\frac{\sin(\pi/p^{\ast})}{2}] \approx-1/\ln(\frac{2p^{\ast}}{\pi})^2$.
It is straightforward to conclude that the asymptotics of ${\cal R}_{(q,p)}$ is solely
governed by the parameter $p$, i.e. ${\cal R}_{(q\gg4,p\gg4)}\to -1/2\ln(\frac{2p}{\pi})$.

\begin{figure}[tb]
{\centering\includegraphics[width=\linewidth]{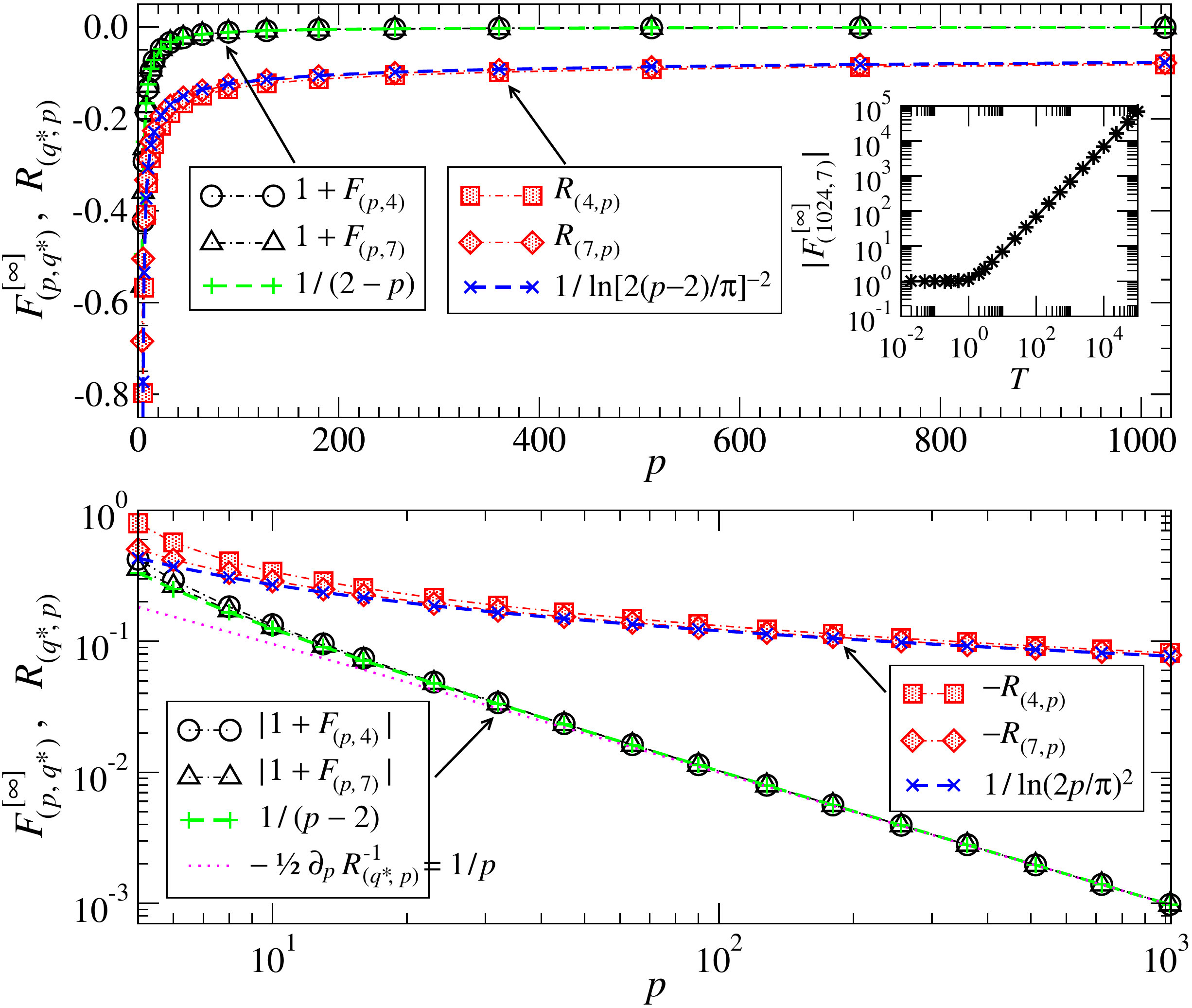}}
  \caption{(Color online) The asymptotic behavior of the free energy per site ${\cal F}
_{(p,q^{\ast})}^{[\infty]}-{\cal F}^{[\infty]}_{(\infty,q^{\ast})}$ for the Ising model
at $T=0.1$ and the Gaussian radius of curvature ${\cal R}_{(q^{\ast},p)}$ for $q^{\ast}=4$
and $q^{\ast}=7$. The asymptotic fitting functions for the energy decay as $-\frac{1}{p}$,
where the ``$+$'' symbols connected with the dashed line fit to the free energy data (the
circles and the triangles). The asymptotics $\propto 1/\ln(2p/\pi)^{-2}$ of the radius
of curvature (the squares and the diamonds) is plotted by the ``$\ast$'' symbols connected
with the dashed line. The inset of the top graphs shows the temperature dependence of
${\cal F}^{[\infty]}_{(\infty,q^{\ast})}$. The log-log plot in the lower graph is used
to enhance the asymptotics of the top graph.}
\label{fig:17}
\end{figure}

Since the seemingly similar $p$-dependence of the free energy plotted on the top graph
in Fig.~\ref{fig:16} does not suffice to conjecture the identical logarithmic asymptotics
as we have derived for ${\cal R}_{(q^{\ast},p\gg4)}$, we extended our numerical calculations
of the free energy per site at larger $p$ for the two selected coordination numbers
$q^{\ast}=4$ and $q^{\ast}=7$. Figure~\ref{fig:17} shows the asymptotic behavior of
${\cal F}_{(p,q^{\ast})}^{[\infty]}$ and ${\cal R}_{(q^{\ast},p)}$ for $4\leq p \leq 1024$
at $T=0.1$. The top and the lower graphs show both the free energy per site and the
radius of curvature in the linear scale and the log-log plot, respectively.

The least-square fitting applied to the free energy per site gives the function
$\frac{1}{2-p}+{\cal F}^{[\infty]}_{(\infty,q^{\ast})}$, which correctly reproduces
the asymptotics of the free energy per site for both $q^{\ast}$.  In contrast to the
radius of curvature, which logarithmically converges to zero as $p\to\infty$, the free
energy per site converges ${\cal F}^{[\infty]}_{(\infty,q^{\ast})}=-1$ for $T\ll1$ and
is linear in temperature ${\cal F}^{[\infty]}_{(\infty,q^{\ast}\lesssim T/2)}=-T\ln(M)$
for $T\gg1$ in accord with Eq.~\eqref{F_inf_T}. At $T=0.1$ the term
${\cal F}^{[\infty]}_{(\infty,q^{\ast})}=-1.00098$ for $q^{\ast}=4$ and $q^{\ast}=7$.
The inset of the top graph in Fig.~\ref{fig:17} shows the functional dependence of the
constant ${\cal F}^{[\infty]}_{(\infty,q^{\ast})}$ on temperature, which is numerically
feasible up to the polygonal size $p=1024$ with the sufficient accuracy (noticing a
negligible dependence on small values of $q^{\ast}$). The log-log plot of the lower graph
clearly demonstrates the difference in the asymptotics ($p\gg4$) between the polynomial
behavior of ${\cal F}_{(p,
q^{\ast})}^{[\infty]}-{\cal F}^{[\infty]}_{(\infty,q^{\ast})} = - \frac{1}{p}$ and the
logarithmic one ${\cal R}_{(q^{\ast},p)}^{-1} = -\ln(2p/\pi)^2$. The thin dotted line
on the lower graph is the numerical derivative of $-\frac{1}{2}{\cal R}_{(q^{\ast},p)}^{-1}$
with respect to $p$, which confirms the convergence to the asymptotic regime $p^{-1}$ of
the free energy per site.
We, therefore, observe an asymptotic relation between the free energy per site and the
radius of curvature on the dual lattice geometry
\begin{equation}
{\cal F}^{[\infty]}_{(p,q)} - {\cal F}^{[\infty]}_{(\infty,q)}
\propto \frac{\partial}{\partial p}{\cal R}^{-1}_{(q,p)}
\approx -\frac{\pi}{2}\exp\left[ \frac{1}{2}{\cal R}_{(q,p)}^{-1} \right]\, ,
\label{scale_dR-F}
\end{equation}
which is valid for any fixed $q\geq4$ and $p\gg4$ (typically $p\gtrsim10^2$) at low
temperatures. The free energy of the multi-spin spin models (or the ground-state energy
for quantum spin systems) on the non-Euclidean lattice geometries exhibits certain
similarities with the Gaussian radius of curvature in the asymptotic regime ($p\gg4$)
we found out as a by-product of our numerical analysis. The necessity of supporting
our findings theoretically is inevitable. The consequences of the current work are
expected to elicit further research, which can bridge the solid-state physics with
the general theory of relativity.

\begin{figure}[tb]
{\centering\includegraphics[width=\linewidth]{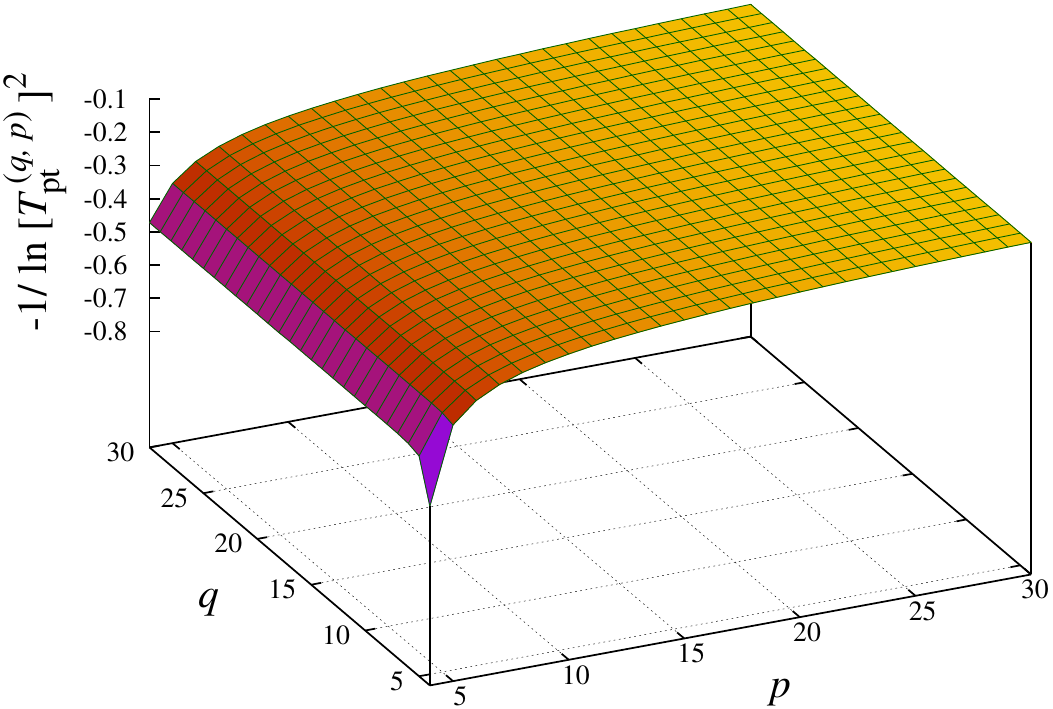}}
  \caption{(Color online) The rescaled phase transition temperatures with respect to $p$
and $q$ are shown in the dual geometry ($q,p$) to emphasize the similarity with the radius
of Gaussian curvature in Fig.~\ref{fig:15}.}
\label{fig:18}
\end{figure}

Having analyzed the phase transition temperatures $T_{\rm pt}^{(p,q)}$ of the Ising model
with respect to the lattice geometries ($p,q$), we again find another analogous relation
for the scaling of the radius of Gaussian curvature
\begin{equation}
-1 / \ln { \left[ T_{\rm pt}^{(p,q)} \right] }^2 \propto {\cal R}_{(p,q)}
\label{scale_dR-T}
\end{equation}
as shown in Fig.~\ref{fig:18}. For better visual comparison with Fig.~\ref{fig:15}, we
plot $-1/[2\ln T_{\rm pt}^{(p,q)}]$ in the dual geometry (i.e., the meanings
of $p$ and $q$ are swapped in the graph). Recall that the higher values of the coordination
number $q$ (for fixed $p$) cause that $T_{\rm pt}^{(p,q)} \propto q$, cf. Eq.~\eqref{Tpt_q},
whereas if $p$ increases (at fixed $q$), the fast convergence to the constant in
Eq.~\eqref{Bethe_q} is achieved.

Hence, the evident mutual similarity of the functional $p,q$-dependence among the free
energy per site (in Figs.~\ref{fig:11}, \ref{fig:12}, and \ref{fig:14}), the radius
of the Gaussian curvature (Fig.~\ref{fig:15}), and the phase transition temperature 
(Fig.~\ref{fig:18}) points to a certain related correspondence, which calls for
consequent theoretical explanations to connect these physical quantities together.

\section{Concluding remarks}

Having generalized the CTMRG algorithm to study the multi-state spin models on an
infinite set of the non-Euclidean geometries ($p\geq4,q\geq4$), we successfully
analyzed the free energy of the models with respect to the underlying lattice
geometry and temperature (including the classification of the phase transitions).
Although such an infinite set of the lattices significantly exceeds those we had
analyzed earlier~\cite{hCTMRGp4,hCTMRG3q}, we have been still facing a challenging
task: to complete the entire set of the regular lattice geometries ($p\geq3,q\geq3$),
which involves all three Euclidean and five spherical geomoetries. To accomplish this
task, a substantially different structure of the recurrence relations of the
($p\geq6,3$) lattice geometries is expected to be derived.

The main purpose of this work was to derive the general formula for the free energy
per spin of the $M$-state clock and Potts models on the ($p,q$) lattice geometries.
The CTMRG algorithm yielded a sufficiently high numerical accuracy with respect to
the exact solutions, which was demonstrated by comparing the phase transition
temperatures obtained by CTMRG and the exactly solvable Ising models on the square
($4,4$) and the Bethe ($\infty,q\geq4$) lattices. Having minimized the free energy
with respect to the lattice geometry ($p\geq4,q\geq4$), we found out that the minimum
of the (bulk) free energy per spin site is satisfied only in the case of the Euclidean
square lattice for each fixed temperature.

Making use of the free energy, the boundary structure of the hyperbolic surfaces
is naturally incorporated into the solution and carries the essential features
of the AdS space. In other words, we have initiated first steps to classify the
properties of regular hyperbolic (AdS) spaces viewed from the condensed-matter physics.
The next step involves a direct numerical calculation of the entanglement entropy of a
subsystem ${\cal A}$ in quantum Heisenberg, $XY$, and Ising models on ($4,q\geq4$) lattice
geometries~\cite{TPVF4q}. Following these steps, we intend to confirm a concept of
the so-called holographic entanglement entropy~\cite{tHooft,Susskind,Takayanagi},
where a non-gravitational theory is expected to live on the boundary of a subsystem
$\partial {\cal A}$ of $(d+1)$-dimensional hyperbolic spaces. The entanglement entropy
$S_{\cal A}$, related to a reduced density matrix of the subsystem ${\cal A}$, can
provide an appropriate measure of the amount of information within the AdS-CFT
correspondence. The entropy $S_{\cal A}$ is proportional to a surface region $\partial
{\cal A}$ (minimal area surface) in the AdS space and is related via duality with the
corresponding $d$-dimensional region ${\cal A}$ defined in CFT. Our consequent study is,
therefore, focused on the analysis of the von Neumann entanglement entropy of quantum
spin systems with respect to the underlying AdS lattice geometry. (In our earlier
studies, we analyzed the entanglement entropy of classical spin systems on hyperbolic 
geometries~\cite{hCTMRG3q,hCTMRG3qn}.)

The surprising inherited physical similarity between the lowest energy of the
microscopic multi-spin models and the radius of the Gaussian curvature certainly
deserves a deeper understanding and theoretical reasoning in the future. Our numerical
observations cannot unambiguously justify our incomplete conjectures of this work. In 
the future, we intend to broaden our current findings to reveal how the intrinsic
structure of the space lattice geometry as well as the microscopic spin-interaction
networks affect the lowest energy, which incorporates information about the entire
system (including the significant influence of boundary effects on the large scales).
Our incomplete conjectures suggest that the free-energy analysis of the simple
multistate spin systems reflects the underlying regular hyperbolic lattice structure
and is proportional to the Gaussian radius of curvature on the dual lattice geometry.

\begin{acknowledgments}
We thank T.~Nishino and F.~Verstraete for valuable discussions.
The support by the projects VEGA-2/0130/15 and QIMABOS APVV-0808-12
is acknowledged.
\end{acknowledgments}

\end{document}